\documentclass[iop]{emulateapj}
\usepackage{graphicx}
\usepackage{amsmath}
\usepackage{ifthen}
\usepackage{afterpage}

\slugcomment{(Draft \today)}

\gdef\gappr{\hbox{$_>\atop{^\sim}$}}
\gdef\lappr{\hbox{$_<\atop{^\sim}$}}

\gdef\Sec{${}^{\prime\prime}$\llap{.}}

\def\etal{\hbox{et al.}}

\gdef\ltsima{$\scriptscriptstyle \; \buildrel < \over \sim \;$}
\gdef\simlt{\lower.3ex\hbox{\ltsima}}

\gdef\gtsima{$\scriptscriptstyle \; \buildrel > \over \sim \;$}
\gdef\simgt{\lower.3ex\hbox{\gtsima}}

\gdef\about{\raise.3ex\hbox{$\scriptscriptstyle \sim $}}

\shortauthors{Kelson \etal\ }
\shorttitle{CSI Redshift Survey: I. Description and Methodology}

\begin{document}

\footnotetext[1]{This paper includes data gathered with the 6.5 meter Magellan Telescopes 
located at Las Campanas Observatory, Chile.}

\title{The Carnegie-Spitzer-IMACS Redshift Survey of Galaxy Evolution since $\lowercase{z}=1.5$:
\break
I. Description and Methodology\footnotemark[1]
}

\footnotetext[2]{Visiting Astronomer, Kitt Peak National
Observatory, National Optical Astronomy Observatory, which is operated
by the Association of Universities for Research in Astronomy (AURA)
under cooperative agreement with the National Science Foundation.}

\author{
Daniel D. Kelson\footnotemark[2],
Rik J. Williams\footnotemark[2],
Alan Dressler\footnotemark[2],
Patrick J. McCarthy\footnotemark[2],
Stephen A. Shectman,
John S. Mulchaey,
Edward V. Villanueva,
Jeffrey D. Crane, \&
Ryan F. Quadri
}
\affil{The Observatories of the Carnegie Institution for
Science, 813 Santa Barbara St., Pasadena, CA 91101}

\begin{abstract}
We describe the Carnegie-Spitzer-IMACS (CSI) Survey, a wide-field, near-IR selected spectrophotometric redshift 
survey with the Inamori Magellan Areal Camera and Spectrograph (IMACS) on Magellan-Baade. By defining a flux-limited sample 
of galaxies in Spitzer \emph{IRAC} $3.6\mu$m imaging of SWIRE fields, the CSI Survey efficiently traces the stellar mass of 
average galaxies to $z\sim 1.5$. This first paper provides an overview of the survey selection, observations,
processing of the photometry and spectrophotometry.
We also describe the processing of the data: new methods of fitting synthetic templates of spectral energy distributions
are used to derive redshifts, stellar masses, emission line luminosities, and coarse information on recent star-formation. Our
unique methodology for analyzing low-dispersion spectra taken with multilayer prisms 
in \emph{IMACS}, combined with panchromatic photometry from the ultraviolet to the IR, has
yielded 37,000 high quality redshifts in our first 5.3 degs$^2$ of the SWIRE XMM-LSS field.
We use three different approaches to estimate our redshift errors and find robust agreement. 
Over the full range of $3.6\mu$m fluxes of our selection, we find
typical uncertainties of $\sigma_z/(1+z)
\la 0.015$. In comparisons with previously published VVDS redshifts, for example, we find a scatter of
$\sigma_z/(1+z) = 0.012$ for galaxies at $0.8\le z\le 1.2$. For galaxies brighter and fainter than $i=23$ mag, 
we find $\sigma_z/(1+z) = 0.009$ and $\sigma_z/(1+z) = 0.025$, respectively.
Notably, our low-dispersion spectroscopy and analysis yields
comparable redshift uncertainties and success rates for both red and blue galaxies, largely
eliminating color-based systematics that can seriously bias observed dependencies of galaxy evolution on
environment.

\end{abstract}



\section{Introduction}
\label{sec:intro}

\noindent Understanding the evolution  of galaxies and large scale structure remains a fundamental challenge 
in astrophysics.  Many ambitious galaxy surveys have been carried out to address this problem, but limited time 
on large telescopes results in a classic problem: sky coverage, depth, and spectral resolution -- choose two.  For 
example, the very-wide-area Sloan Digital Sky Survey (SDSS)  provides a wealth of spectral information for 
galaxies over a cosmologically significant volume, but its modest depth limits the SDSS to the relatively local, 
`modern' universe.  Conversely, the Hubble Ultra-Deep Field photometric survey probes deep into cosmic 
time --- to the epoch of reionization, but only over a tiny volume whose small population of galaxies leads to large 
uncertainties on their physical properties. 
While the union of such surveys has begun to paint a coherent picture 
of galaxy growth, significant patches of blank canvas limit our ability to fully describe
how environmental processes, the infall of 
gas that fuels star formation, galaxy mergers and acquisitions, and feedback, 
all shape the evolution of galaxies over cosmic time.

The Carnegie-Spitzer-IMACS Survey (CSI) has been designed to address one of the most dramatic and least
understood features of galaxy evolution --- the remarkably rapid decline in cosmic star formation since
$z\sim 1.5$.  It is during this extended epoch of galaxy maturation that galaxy groups and clusters have also
emerged as a conspicuous feature of the landscape.  The CSI Survey is uniquely able to link together the 
evolution of individual galaxies with these features of large-scale structure growth.

In our ambitious spectrophotometric redshift survey of distant galaxies, we strike a balance between
the aforementioned three factors: (1) high completeness to moderate redshift ($z\simlt 1.5$), 
(2) spectral resolution intermediate between conventional photometric and spectroscopic 
surveys (combining the efficiency of imaging surveys with a spectral resolution high enough to 
resolve large-scale structure and prominent emission-lines); and (3)  an unprecedented 
area of 15 deg$^2$ for a $z\simgt 1$ survey.  This gives the CSI Survey a volume
comparable to the SDSS, with a selection method that efficiently traces stellar mass over 2/3 
the age of the universe ($0.4<z<1.5$) --- spanning the critical redshift range where cosmic
star formation precipitously drops, and groups and clusters become prominent.  
As the redshift survey with the largest unbiased volume at $z=1$,
CSI will allow us to
comprehensively address the interplay between environment, galaxy mass buildup, and star formation
at these redshifts.

In this, the first paper of the CSI Survey, we describe the design of the project, with particular focus 
on the flux limits of the selection, the data processing and the spectral energy distribution (SED) fitting 
with its attendant determination of redshifts and redshift errors. We apply our methodologies to the first 
batch of data in 5 degs$^2$ of the SWIRE XMM-LSS field, and provide an overview of some basic 
properties of galaxies over the past 9 Gyr, with an emphasis on the gains made by selecting galaxies by 
stellar mass instead of by their rest-frame UV light.  

\begin{figure*}[htb]
\centerline{
\includegraphics[width=5.8in]{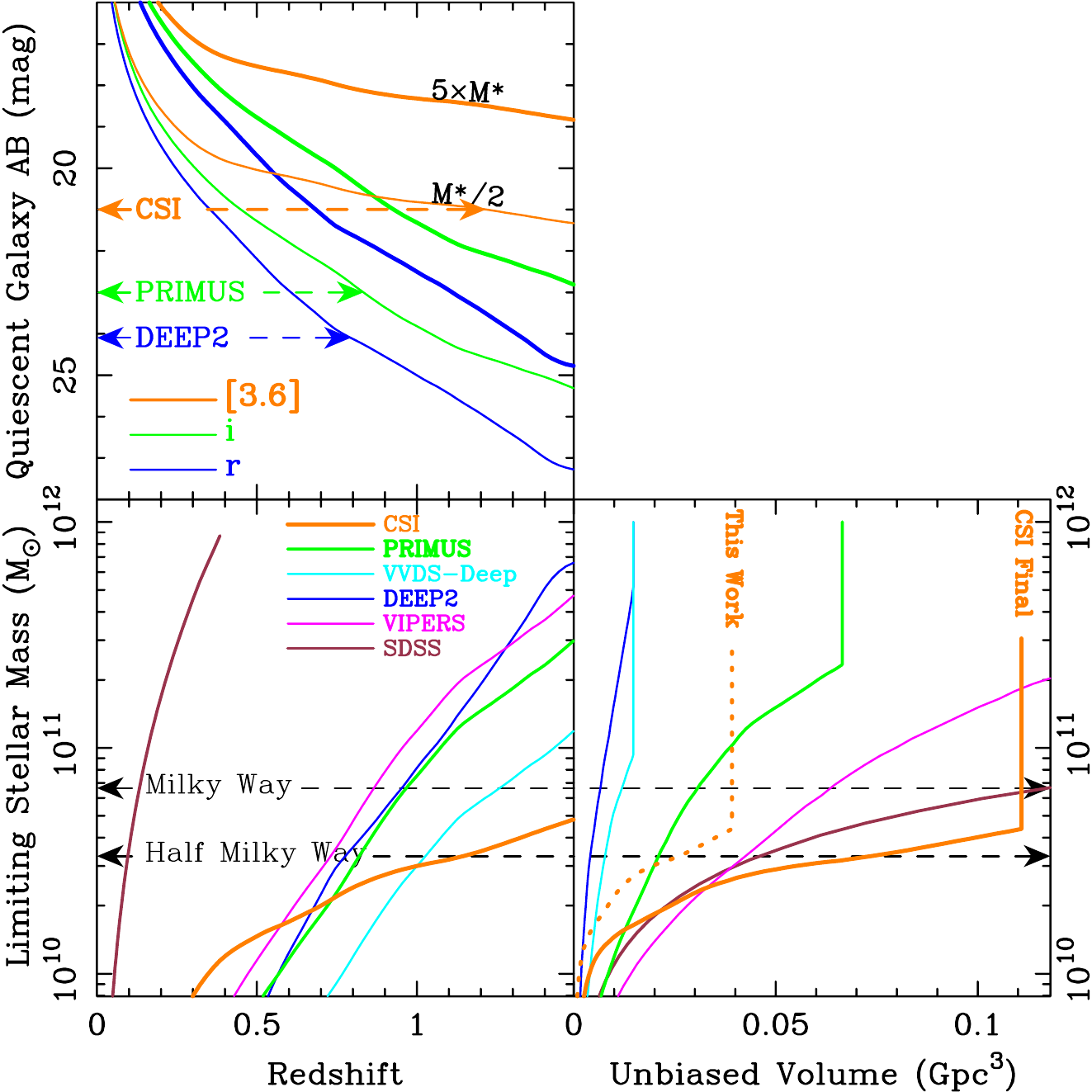}
}
\caption{(top left) Apparent magnitudes in $r$, $i$, and $3.6\mu$m as a
function of redshift, for passively evolving stellar populations ($3 \le z_f\le 6$) with
stellar masses $5M^*$ and $M^*/2$ \citep[$\log M^* \sim 10.85$ at $z \lesssim 1$;][]{drory2009}.
The magnitude limits for DEEP2 \citep{willmer2006} and PRIMUS \citep{coil2010} are
drawn. Such optical flux limits only cover the most massive passively evolving systems or those
galaxies with young unattenuated stellar populations, biasing galaxy samples at early times. The shallow
dependence of the $3.6\mu$m magnitude on redshift yields a selection with significantly less bias
against old systems.
(bottom left) Limiting stellar mass of faint galaxy surveys by redshift. By using the \emph{IRAC}
$3.6\mu$m band, the CSI survey (the solid orange line) traces stellar mass more uniformly than
samples selected in the optical.  Our sample reaches stellar
masses equivalent to the present day Milky Way
\citep[$6.6\times 10^{10}M_\odot$;][]{mcmillan2011} out to $z=1.4$, almost order-of-magnitude lower than
DEEP2, and half the present day Milky Way at $z=0.9$, a factor of two deeper than DEEP2 or PRIMUS.
(bottom right) Volumes probed with complete, unbiased samples for several redshift surveys as functions of limiting stellar
mass. When the 15 deg$^2$ is completed, CSI's volume will be more than an order of magnitude
larger than DEEP2. The volume traced by
the first 5.3 degs$^2$ (this paper) is shown by the dotted line.
\label{fig:masslimits}}
\end{figure*}


\section{The Carnegie-Spitzer-IMACS Survey}
\label{sec:survey}

\subsection{The limitations of optically-selected surveys}

Any effective probe of galaxy assembly must sample a wide range of masses in order to trace 
evolutionary connections between large and small systems. The build-up of massive red sequence 
galaxies may be driven by mergers with sub-$M^*$ systems, so it is essential to trace evolution to 
masses below $M^*$ to mitigate against the differential growth of the high- and low-mass 
populations. This trade-off between depth and area noted above has led to a dichotomy in 
redshift surveys. Very deep programs, such as the Gemini Deep Survey \citep[GDDS;][]{abraham2004} or
the Galaxy Mass Assembly Ultra-deep Spectroscopic Survey \citep[GMASS;][]{halliday2008}
could not cover enough volume to robustly sample the evolution of the high mass population at 
$z < 1$, while other surveys that have traded depth for area do not reach below $M^*$ with high 
fidelity. 

An old stellar population at $z=1$ with a stellar mass of $10^{11}M_\odot$
corresponds to roughly $i=23$ mag and $r=24$ mag in the optical, as shown in Figure
\ref{fig:masslimits}(top left).  The extreme optical faintness of such 
galaxies results in many of them being missed in optically-selected surveys despite
their relatively high masses. For example, the
selection limits for DEEP2 \citep{willmer2006} and PRIMUS \citep{coil2010}
are shown in this figure by the blue and green dashed lines, respectively. At $z>1$
these programs {\it required\/} galaxies to be extremely massive or have their optical light dominated
by young stellar populations in order to be fall within the survey selection.
At best it can be difficult to trace the evolution of the most massive galaxies with
such surveys. At worst, if not properly accounted for, color-dependent selection effects
introduce biases and systematic effects. Extremely deep optical limits can be taken as one valid approach to ameliorating this
problem, with the side effect of an overwhelming number of low-mass star-forming dwarfs dominating one's source catalog.

\subsection{The potential of an \emph{IRAC}-selected survey}

The integrated light of all but the youngest stellar populations are dominated by light from 
the stellar giant branch, cool stars whose light output peaks in the near-IR.  It has long been
recognized (e.g., Wright \etal\ 1994) that this results in a $1.6\mu$m ``bump'' -- a peak in 
bolometric luminosity -- for galaxies with a wide range of star formation histories.
Indeed, as shown clearly in (for example) Figure 1 of \cite{sorba2010}, this feature is nearly unchanging in 
\cite{bc2003} models of stellar populations with mean ages 100 Myr $< \tau < $10 Gyr.  
Put in other terms, the near-IR mass-to-light ratio, M/L$_{1.6\mu m}$ changes slowly for all but the 
youngest stellar populations.  

The CSI Survey exploits the wide field and sensitivity of the \emph{Spitzer Space Telescope} 
with the \emph{Infrared Array Camera} (\emph{IRAC})  to take advantage of this property of the integrated near-IR light 
from stellar populations in $z\sim 1$ galaxies.  Combined with the insensitivity to internal and Galactic
extinction, selection at $3.6\mu$m closely mirrors selection by stellar mass.  Figure 1
(top left) directly compares the evolution of $3.6\mu$m magnitude with the optical $r$ and $i$
bands: the CSI selection wavelength has a dependence on redshift that is much shallower than surveys
selected in the optical. What slope remains for the CSI selection function is the unavoidable
\emph{k-correction}: over the redshift range $0.7 < z < 1.5$, the center of the $3.6\mu$m
\emph{IRAC} band
corresponds to restframe wavelengths of $2.1\mu m > \lambda_c > 1.4\mu m$, that is, straddling the
$1.6\mu$m ``bump'' of the spectral energy distribution (SED), but shifting through this feature as
the redshift changes.  

Even so, massive galaxies exhibit a much flatter trend of observed magnitudes with redshift
in the IR than in the optical, and this weaker dependence on galaxy mass afforded by $3.6\mu$m-selection of the sample is a 
key feature of the CSI Survey.
Our goal has been to make a spectrophotometric survey to characterize the galaxy populations and 
environments to $z=1.5$, unbiased at all redshifts down to a stellar mass of $M = 4 \times 10^{10} M_\odot$.
As can be read from Figure 1, this mass limit corresponds at $z=1$ to $r=25$ mag or $i=24$ mag, with 
equivalent limits of $r=26$ mag and $i=25$ mag to reach this mass limit at $z=1.2$.
Our current spectroscopic reduction and analysis, described below, is reaching
an effective photometric limit of $r=26$ mag (the dashed orange line).

Figure 1 (bottom left) plots the limiting mass as a function of redshift for CSI (solid orange)
and other redshift surveys. The depth of CSI in stellar mass is
substantially less sensitive to redshift compared to the others due to the \emph{IRAC} $3.6\mu$m selection,
varying by a factor $\sim 3$ over $z=0.5-1.5$, compared to $1-2$\,dex over the same redshift
range for optically-selected surveys. The result
is that CSI samples nearly uniformly by stellar mass for the full, large volume of the survey,
and to a lookback time of 9 Gyr.

This critical point is illustrated in Figure 1 (bottom right),
plotting the depth in stellar mass against the volume probed with complete, unbiased samples.
DEEP2, shown in violet (the line colors are the same as in the bottom left), is limited in both area
and mass depth. PRIMUS's 9 degs$^2$ is limited in depth, and thus only probes an
unbiased volume comparable to our first 5 degs$^2$. When CSI reaches its goal of 15 degs$^2$, the
survey will cover an unbiased volume equal to the SDSS with similar depth in stellar mass.
The large areas available from legacy Spitzer \emph{IRAC} surveys --- both wide {\it and} deep, and the
freedom from foreground and internal extinction at this wavelength, 
allows the construction of uniformly deep, homogeneous photometric samples for spectroscopic follow-up.

CSI reaches factors of 2-6 deeper than DEEP2 in mass, over an area
ultimately 8 times wider, and redshifts sufficiently accurate to characterize
environments by directly identifying groups and clusters. With such data we
aim to make the first group catalog at $z=1$ that is statistically comparable to SDSS,
and thus enable the first detailed environmental characterizations of galaxies at a time when the
universe was less than half its current age.

\begin{figure*}[htb]
\centerline{
\includegraphics[width=6.0in]{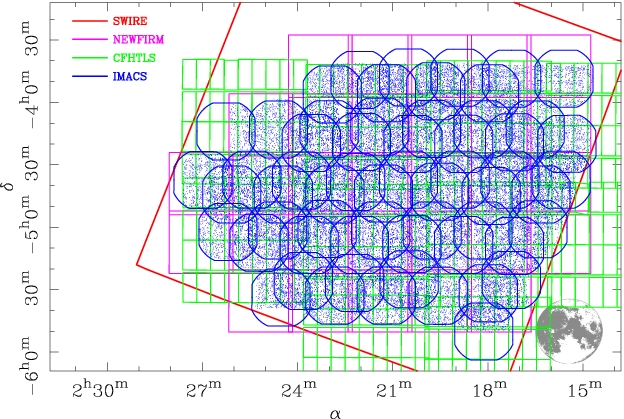}
}
\caption{A schematic of observations in the first 5 degs$^2$ of the SWIRE-XMM field
studied by the Carnegie-Spitzer-IMACS survey. The SWIRE \emph{IRAC} imaging field is outlined in
red. The CFHTLS-W1 optical data reanalyzed by us is shown in green. The first set of
NEWFIRM $J$ and $K_s$ observations are marked in violet. The positions of the IMACS
slitmasks are outlined in blue, with the blue points marking the positions of the 37,000
galaxies with CSI redshifts. The field of view of IMACS is comparable to the size of the
full moon, illustrated in the lower right.
\label{fig:xmmfield}}
\end{figure*}

In this first paper, we describe the details of our data reduction and redshift
fitting, redshifts for the first 37,000 galaxies in 5.3 degs$^2$ of the
SWIRE-XMM field (Figure \ref{fig:xmmfield}), and an initial characterization of the general galaxy populations
in our stellar mass-limited sample.

\section{Data}
\label{sec:data}

A project of this type and scope requires attention to detail and care in the processing of a broad range of
data from different sources. In this section we describe the imaging data and reductions that underpin the 
broadband flux measurements for the sample, both for the ultimate goal of fitting SEDs, but also for
defining the selection criteria and incompleteness functions. Unless otherwise specified, all object 
detection was performed using SExtractor \citep{bertin1996}.

\subsection{Photometry}
\label{subsec:photometry}

The SWIRE Legacy Survey \citep{lonsdale2003} observed several large fields with \emph{IRAC} to a depth
suitable for our survey. Three of these fields are accessible from southern telescopes, providing up
to 23.8 deg$^2$ of 
coverage at $3.6\mu$m. Basic properties of the three fields are given in Table \ref{tab:fields}.
Of these fields, $15.3$ deg$^{2}$ have supplemental optical imaging
that is publicly available.
Of the 9.1 degs$^2$ of \emph{IRAC} imaging in the XMM-LSS field,
6.9 degs$^2$ is covered by the CFHT Legacy Survey
W1 $ugriz$ imaging. In this section we discuss our analysis of the SWIRE XMM-LSS \emph{IRAC}
data, our reprocessing of the CFHTLSW1 $ugriz$ imaging and subsequent  photometry of the $3.6\mu$m
catalog, as well  as our observations at $J$ and $K_s$ of the field using NEWFIRM \citep{autry2003}
on the Mayall 4m telescope at Kitt Peak National Observatory.

\subsubsection{Spitzer-\emph{IRAC} Imaging}
\label{subsubsec:IRAC}

The SWIRE Legacy Survey was a program undertaken to trace galaxies by stellar mass back to
$z=2$ \citep{lonsdale2003}.
The \emph{IRAC} images of the XMM-LSS field were obtained from the archive at IPAC. The reductions
and processing of these data were described by \cite{surace2005}. In order to 
minimize contamination of the object catalog by artifacts around stars, we configured SExtractor so as to ignore elliptical 
regions around bright stars. A `mexican hat' convolution kernel was used for object detection, resulting in 585,159 
objects in the $3.6\mu$m catalog of the XMM-LSS field. Fluxes were measured in a manner described by
\cite{surace2005}. Down to our selection limit of $3.6_{AB}=21$ mag the
$3.6\mu$m catalog contained 266,621 objects.

\begin{deluxetable}{l c c l}
\tablecaption{Basic Properties of the Southern SWIRE Fields Targeted by CSI
\label{tab:fields}}
\tablehead{
\colhead{Field} &
\colhead{SWIRE Area} &
\colhead{SWIRE+Optical} &
\colhead{Filters} \\
&
\colhead{(degs$^2$)} &
\colhead{(degs$^2$)} \\
}
\startdata
XMM-LSS & 9.1 & 6.9 & ugriz\\
ELAIS S1 & 6.9 & 3.6 & BVRIz \\
CDFS & 7.8 & 4.8 & ugriz
\enddata
\end{deluxetable}

\subsubsection{The Optical Imaging}
\label{subsubsec:optical}

The CFHT Legacy Survey \citep{cuillandre2006} targeted multiple fields using the one-degree MegaCam imager around 
the sky with a broad range of scientific goals. The ``Wide'' survey focused on several astrophysical questions, with the W1 
field providing almost 7 deg$^{2}$ of overlapping $ugriz$ coverage in the SWIRE XMM-LSS field. Unfortunately, the processed 
images and catalogs available at Terapix were not entirely suitable for our purposes, owing to uncertain astrometry and 
improper defringing of the $i$ and $z$ data. Therefore, we obtained the complete set of calibrated frames from the CFHT 
archive and processed the data using the following additional steps. First, we constructed new fringe frames in $i$ and 
$z$ by medianing scaled exposures obtained within a single night. These were then rescaled and subtracted from the 
individual exposures. Some bright objects did bias these medians when the number of exposures was small (i.e. $N \lappr 10$), 
leaving faint traces in the resulting fringe frames.  In general, however, the greater sky uniformity after subtraction of these new 
fringe frames improved the depth of our resulting catalogs by $\sim 0.2$ to $0.5$ mag compared to the on-line catalogs. We 
constructed sky frames as well for the $u, g$, and $r$ bands using a similar methodology, though not restricting the 
construction to data collected within single nights.

New astrometric solutions were derived for each exposure using the \emph{IRAC} catalog as a set of deep astrometric standards. 
Cubic solutions for each chip were derived with a typical RMS scatter of 0\Sec 15. For those regions of the MegaCam images that
did not overlap the SWIRE data, we supplemented the catalog with the 2MASS point source catalog \citep{skrutskie2006}, in order 
to prevent the solutions from diverging at the edges of the \emph{IRAC} field.

Using these new astrometric solutions, and the photometric solutions from the headers of the individual frames, we constructed 
cosmic-ray-cleaned image stacks with 0\Sec 185 pixels, 30 arcmin on a side and evenly spaced. These smaller images were more easily 
managed than the larger image formats provided by Terapix, which allowed us to distribute the image analysis and photometry 
tasks over multiple processors. The zeropoints were checked by comparing the photometry of moderately bright objects with the 
data in the catalogs from Terapix and we found systematic offsets less than $\pm 0.03$ mag in every case.
Object catalogs were
generated using SExtractor, and these were matched to the \emph{IRAC} catalog with a tolerance of
1\Sec 5 arcsec, with the goal
of using the optical image characteristics to aid in star/galaxy separation, and to determine which \emph{IRAC}-selected objects 
have centroids with optical centers falling far from the defined slit positions.

\subsubsection{The Near-IR Imaging}
\label{subsubsec:nearir}

NEWFIRM \citep{autry2003} is a wide-field ($27^\prime \times 27^\prime$) near-IR camera deployed by NOAO at the Mayall 4m at Kitt Peak from 2007 to 
early 2010. During the fall semesters we imaged the XMM-LSS field in $J$ and $K_s$. Typical exposure times were
70 min/pixel in $J$ and 32 min/pixel in $K_s$. The seeing ranged between $0.8 - 1.4$ arcsec (FWHM).

The data were processed using a fully automated, custom pipeline, written as a prototype for a wide-field imager being deployed at
Magellan. The basic steps in the reduction were:
(1) subtraction of the dark current;
(2) correction for nonlinearity;
(3) division by a flat-field;
(4) masking of known bad pixels;
(5) construction of first-pass sky frames;
(6) derivation of image shifts, in arcsec, on the sky;
(7) stacking of the first-pass sky-subtracted frames;
(8) generation of a deep object catalog;
(9) construction of object and persistence masks for second-pass sky estimation;
(10) temporary interpolation over objects, persistence, and bad pixels for each frame;
(11) constructing bivariate wavelet transforms of these masked frames;
(12) fitting the time variation of the wavelet transforms of the sky using
shorter frequencies for larger spatial scales;
(13) reconstruction of model sky frames by inverting the temporally smoothed wavelet transforms;
(14) subtraction of the model sky frames;
(15) re-derivation of image shifts, in arcsec, on the sky;
(16) final stacking of the sky-subtracted frames, including the generation of sigma and exposure maps;
(17) identification of objects in the 2MASS point source catalog, restricted by 2MASS PSC quality flags;
(18) application of rotation, translation, and scale to the camera distortion based on the 2MASS objects; and
(19) calculation of image zero-points using the 2MASS objects.

Using the zeropoints of each image, we constructed image mosaics $30'$ on a side centered at the locations of
our $ugriz$ mosaics, but with 0\Sec 4 pixels.

\subsection{Aperture Photometry}

Magnitudes in $ugrizJK_s$ were derived using aperture photometry within a range of circular apertures for the
entire SWIRE catalog.  Because the seeing varied with wavelength and pointing, we convolved the $z$-band 
image stacks with Gaussians to simulate the effects of poorer seeing in the $ugriJK_s$ data. The offsets in 
magnitude from the degraded $z$ images were applied as PSF-corrections to the $ugriJK_s$ aperture magnitudes. 
While the details of the PSF and potential blending are critical for small apertures and for modeling the profiles of 
the objects, our choice of fitting the SEDs to much larger-aperture ($D=4"$) magnitudes allows us to utilize a simple and economical 
approach to the PSF corrections. This choice was especially important given the enormous size of the object catalog, 
while reducing the systematic errors in matching the aperture magnitudes to the \emph{IRAC} $3.6\mu$m fluxes.

\begin{figure}
\centerline{
\includegraphics[width=3.0in]{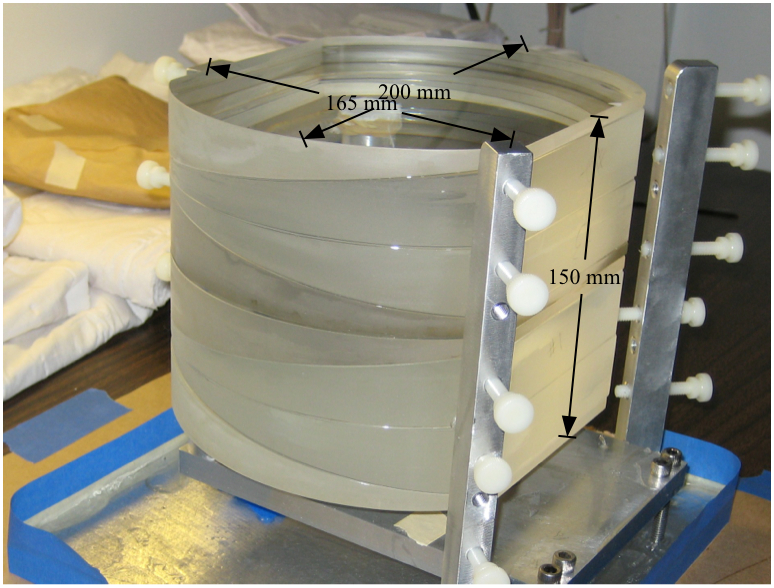}
}
\caption{Construction of the new ``Uniform Dispersion Prism,'' built for CSI.
Dr. S. Shectman designed the UDP to have a resolution of $R\sim 25$
from 7500\AA\ to 1$\mu$m. The eight layers are made from thin prisms of S-FPL51 and N-KZFS2, to make a stack
of glass 150 mm thick. This prism has a resolution with a mild dependence on wavelength (see Figure \ref{fig:resolution})
compared to the first prism deployed in IMACS \citep[see][]{coil2010}.
\label{fig:udpphoto}}
\end{figure}

\begin{figure*}[t]
\centerline{
\includegraphics[width=7.0in]{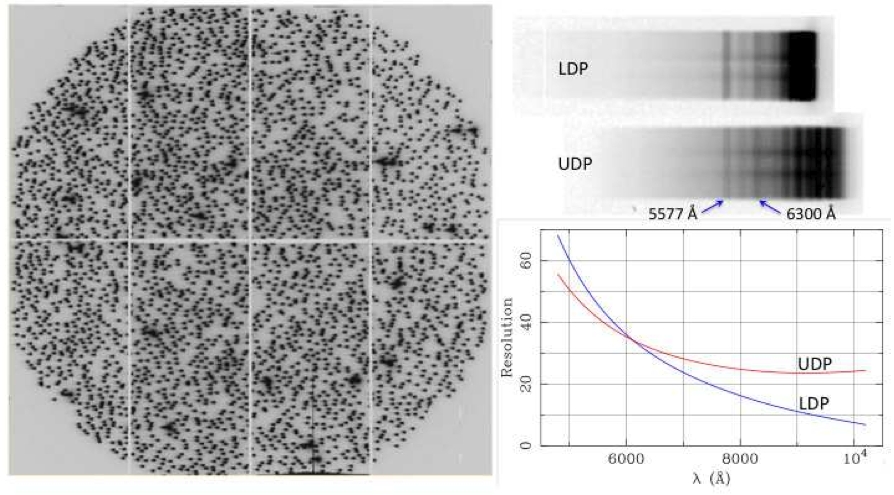}
}
\caption{\citep[Figure reproduced from][]{dressler2011}
(left) Multi-slit mask designed for use with the Low Dispersion Prism.
(top right) Example LDP spectrum --- nod \& shuffle \citep{glazebrook2001} produces
doubled object and sky spectra. Night sky lines [O I]5577\AA\ and [O I]6300\AA\ are
marked. Below the LDP spectrum is one obtained with the UDP, with its more uniform
dispersion (bottom right).
The resulting difference in resolution is shown in the bottom right. The resolution
provided by the new Uniform Dispersion Prism, designed for CSI, provides an improved
ability to trace the 4000\AA\ break and [OII]3727\AA\ to $z=1.4$ when coupled with IMACS's
red sensitive e2v detectors.
\label{fig:resolution}}
\end{figure*}

\subsection{Spectroscopy}
\label{subsec:spectroscopy}

Between 2008 and 2009 the SWIRE XMM field was targeted in 31 multislit mask exposures with the \emph{IMACS} 
f/2 camera configuration and the Low Dispersion Prism (LDP) designed by S. Burles for the PRIMUS redshift survey
\citep{coil2010}. In 2010 we observed 29 SWIRE XMM masks using an innovative eight-layer disperser called the
Uniform Dispersion Prism (UDP, see Figure \ref{fig:udpphoto}).
Although both of these dispersers produce spectra with median resolutions of $\lambda/\Delta\lambda \sim 30$, 
the dispersion curve of the UDP is much flatter, providing significantly higher resolution out to $1\mu$m than the LDP at the
expense of marginally lower resolution in the blue (Figure \ref{fig:resolution}).

Previous work by the PI on a similar program had obtained adequate $S/N$ ratios with exposure times of 3 hours 
down to $z'=23.3$ mag using less sensitive SITe detectors (Patel et al. 2009ab, 2010, 2011). Significantly more 
sensitive E2V detectors were installed in IMACS, thanks to support from the NSF's TSIP program (see Dressler 
\etal\ 2011); these CCDs boosted the throughput of the instrument by factors of 2-3 in the far red. We observed 
with these detectors using the Nod and Shuffle mode of IMACS
\citep[see][]{glazebrook2001}, and accumulated exposure times of 2h per galaxy for more than 90\% of
the sample, using individual integrations with durations of 30 min (15 min per position).
Typical slit lengths were 5 arcsec. The positions of the slit masks are shown in Figure
\ref{fig:xmmfield}.

Spectrophotometric standards were obtained during most observing runs, with a consistency of 5\% from run-to-run. Helium 
calibration lamps illuminating the deployable flat-field screen were taken approximately every hour (i.e., every other science 
exposure). In this section we describe the basic steps involved in reducing the multislit prism data using the custom pipeline 
that had been written for the earlier cluster programs
\citep[e.g.][]{patel2009a,patel2009b,patel2011}.

\subsubsection{Wavelength Calibration}

Two critical aspects of the reduction of low-dispersion prism data involve mapping wavelengths on the detector,
and transposing between sky coordinates of objects and CCD coordinates.
Too few calibration lamp lines can be used to define the solution for a given slitlet, so we derive
global wavelength mappings over an entire CCD using all of the slitlets simultaneously. Thus, lines
in a single spectrum that may be corrupted by bad columns or pixels will not affect the solution for
that slitlet, as the $\sim 3000$ helium lines over a CCD frame constrain the fit.

The following are the basic, automated steps we have implemented in our pipeline: 
\begin{itemize}
\item Derive mapping of sky coordinates to CCD coordinates using an image of the slitmask;
\item Identify lines and fit 2D wavelength solutions using isolated helium lamp exposures on a chip-by-chip basis;
\item Refit new centroids for blended helium lines using nonlinear simultaneous Gaussian fitting on a slit-by-slit basis;
\item Fit for improved mapping of 2D wavelength solutions;
\item Shift the wavelength maps using 2-3 night sky emission lines (e.g., [O I] 5577\AA, Na 5890\AA, [O I] 6300\AA);
\end{itemize}

We start with a
distortion map for the camera, defined from previous exposures of the field around the globular cluster Palomar 5, to compute 
the approximate mapping of sky coordinates to CCD coordinates for the objects in the multislit mask. SExtractor is run on a 
direct image of the slit mask and a simple pattern analysis matches up the predicted positions of the slits and the measured
positions of the slits to create an adjusted mapping of sky coordinates to CCD coordinates.

With a theoretical
dispersion curve for the prism, we generate a trivariate mapping of sky coordinates to CCD coordinates that is wavelength 
dependent. SExtractor is run on a helium comparison lamp image and the resulting catalog of the 3,000 to 4,000 helium lines 
is pattern-matched to the predicted positions of 8 unblended helium lines (out of the 11 lines in our line list). New wavelength 
mappings are solved for as the order of the fit is gradually increased. The nearly final solutions generated at this stage are 
first order rescalings of the theoretical dispersion, with coefficients that are cubic (check!) polynomials of the sky coordinates. 
The typical RMS at this stage is 0.4 pixels (check). With these solutions, we can now generate predicted positions for the 
complete helium line list and fit Gaussian profiles to the full line list for each slit. Within each slit, the Gaussians are fit 
simultaneously, providing accurate positions for both the unblended and blended helium lines. Using these new positions for 
the helium lines in CCD coordinates, we derive through iteration, the trivariate mapping between CCD coordinates and
both wavelength and sky positions, with a typical final RMS scatter of 0.25 pixels. For LDP data this scatter corresponds 
to $\sim 50\,$\AA at  $9000\,$\AA, or $\delta z \sim 0.005$. This large uncertainty in the wavelength calibration in the far red is 
due to the fact that the dispersion reaches $\sim 200\,$\AA/pixel at $9000\,$\AA; the scatter of $\sim50\,$\AA\ introduces a 
25\% uncertainty in $\Delta\lambda$ at such wavelengths when computing the flux calibration. Because the resolution with the 
UDP is $3\times$ higher at these red wavelengths, this additional source of uncertainty is negligible for those data.

The final step involves cross-correlating simulated night sky emission lines ([OI]5577\,\AA, Na I 5890\,\AA, and [OI]6300\,\AA)  
placed at their predicted positions with the spectra in the science frames, on a slit-by-slit basis. The median offset in both $x$ 
and $y$ is then applied to the wavelength mappings found from the nearest helium lamp exposure.

\subsubsection{Extracting Spectra}

Due to the unique nature of these prism data, the science frames must also be
processed in a nonstandard fashion using custom written routines. We take particular care in the 
extraction of spectra because of the small number of pixels covered by each object, the faint optical limits we expect to probe, 
and the proximity of the objects to slit ends where residual sky counts can bias simple extraction algorithms.
Starting with the two 2-dimensional spectra (here denoted ``A'' and ``B'') of each galaxy that come from the nod-and-shuffle observations,
the basic processing steps are as follows:

\begin{itemize}
\item Subtract spectrum A from B, and B from A;
\item Divide by a normalized flat-field, using pixels only covered by spectrum A;
\item Optimally extract each object within a given exposure, adopting a wavelength-dependent Gauss-Hermite
expansion of the spatial profiles for each object, and using the pixels of the A and B spectra simultaneously;
\item Solve for the global translation, rotation, and scale changes in the object positions due to slitmask misalignment;
\item Fix the object positions and redo the optimal extractions, deriving new Gauss-Hermite moments and 1D spectra.
\item Combine the data from all exposures of an object into a single extracted spectrum.
\end{itemize}

A more detailed description of these steps follows.
Initially we subtract the night sky background spectrum, using the ``B'' spectrum
as the sky for the ``A'' spectrum, and vice versa. Next we divide the results by a
normalized flat-field using the pixel locations of the ``A'' spectrum for
both ``A'' and ``B'' spectra (reminder to the reader unfamiliar with the details of Nod \& Shuffle:
electrons counted in the ``B'' pixels originated in the ``A'' pixels).

Recovering the one-dimensional spectra of the objects in the slitlets is now possible, but several additional steps and
some care are required, because of the proximity of the objects to the ends of the slitlets, and given the relatively short 
nods ($1.6$ arcsec). Our procedure is a modification of the standard optimal extraction \citep{horne1986} with several 
key differences. Crucially we can use both the ``A'' and ``B'' spectra simultaneously to measure the spatial profile of a 
given object. We employ a Gauss-Hermite  decomposition of the spatial profile, with moments that are low-order 
Legendre polynomial functions of wavelength. An iterative routine also solves for the centroid of the Gaussian as a 
low-order polynomial of wavelength. The second moment (e.g., standard deviation of the Gaussian) is also determined 
as yet another low-order function of wavelength. A first guess for the one-dimensional spectrum of the object can be obtained 
using this Gaussian approximation for the spatial profile, by solving for the b-spline (Dierckx 1983) that fits the ``A'' and 
``B'' data, both normalized by the wavelength-dependent Gaussian spatial profile. With this approximate spectrum, we 
resolve for the parameters of the spatial profile, including up to 4 Hermite moments (10 for spectrophotometric flux standards), 
each also wavelength-dependent. The wavelength binning of each slitlet is preserved, defined by the wavelength solution
at its location.

With a first set of extracted spectra and, more importantly, object moments, in all eight CCDs of a given exposure, we analyze 
the difference between the predicted object positions and the detected object positions. Deviations are well described by a linear 
function of coordinates in the slitmask, largely the result of small misalignments when the observations were taken. We use this 
new linear function to fix the object positions, re-compute spatial profiles, and re-extract object spectra.

Using the spatial profiles of each observation of a given object, we have a final procedure that solves for the one-dimensional 
spectrum using all exposures simultaneously, using the flux calibration (see below) so that all data are now in physical units. 
We fix the first and second moments to those determined above, and employ a similar optimal extraction routine to that described 
earlier, but one which uses the ``A'' and ``B'' spectra of all the exposures, recomputing the higher-order Hermite moments of the 
spatial profiles and solving for a b-spline that best represents the spectrum of the object in a least-squares sense. Pixels are weighted 
according to the inverse of the expected noise. Bad columns and flagged pixels are given zero weight. Iteration allows us to flag and 
reject most cosmic rays. The process performs an optimal extraction that yields a single 1D spectrum that matches all the data in a 
least-squares sense. Given the small differences between the wavelength solutions of the exposures, we sample the one-dimensional 
b-spline representation of the spectrum at a fixed set of wavelengths, defined by the dispersion of the prism at the center of the field. 
This fixed grid greatly simplifies the construction of templates for fitting the SEDs.

Spectra for objects repeated in overlapping slitmasks were not combined at this stage, but kept separate --- for this paper --- so that we could 
better assess data quality and empirically estimate our redshift precision. Approximately 10\% of the galaxies were repeated; we discuss 
these below in the context of the redshifts and their uncertainties. Approximately 20\% of the objects observed with the LDP were also 
reobserved with the UDP.

\subsubsection{Spectrophotometric Flux Calibration}

Accurate flux calibration of the spectra is a critical component of the survey because the relative slope of a spectrum provides useful 
information for the determination of redshift. For CSI we choose bright hydrogen white dwarf standard stars with deep,
broad Balmer lines that can be detected with the LDP/UDP. Deep Balmer lines are required because we expose these stars in
the alignment star boxes and the positions of the absorption lines allow us to refine the zeropoints
of the wavelength calibrations for the stars when they are not accurately centered in the boxes.

The processing of standard star exposures is similar to that described above, with the following modifications. First, to capture 
essentially all of the photons from the star, we place it in the center of an alignment star box on one of our slitmasks. By doing so we 
can infer the wavelength calibration for the alignment star box by using the trivariate wavelength solution of the helium lamp exposure
obtained immediately after the standard star.

In order to derive the sensitivity function, we must first compensate for the fact that the seeing in the standard star exposure leads to 
different resolution than one would obtain with a narrow slit. We make an assumption that the seeing was isotropic and that the Gauss-Hermite 
parameterization of the spatial profile is a valid descriptor of the seeing profile in the dispersion direction. Using this profile as a kernel, we 
deconvolve the extracted spectrum using an implementation of {\sc CLEAN\/} \citep{hogbom1974}, convolving the result to a resolution 
equivalent to that defined by the science data (FWHM=4.5 pixels).

Due to time and weather constraints, spectrophotometric flux standards were not obtained during every night, but the excellent run-to-run repeatability of the relative 
flux calibration from 8500\,\AA\ to 4500\,\AA\ for the LDP and from 9000\,\AA\ to 4500\,\AA\ for the UDP, of approximately $\pm5\%$
allows us to use flux standards obtained on different nights or during different runs.

\begin{figure*}[t]
\centerline{
\includegraphics[width=7in]{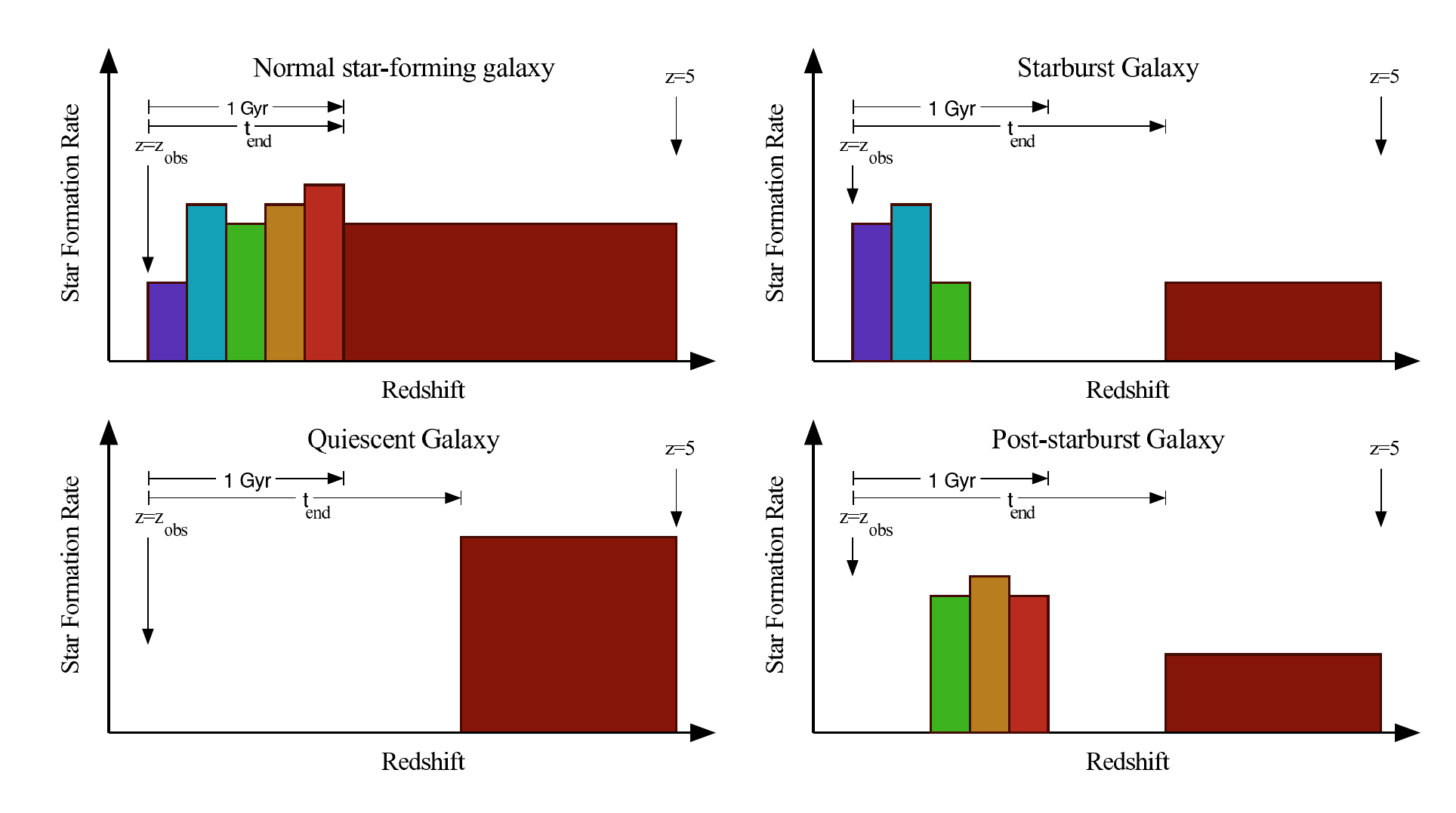}
}
\caption{Cartoon star formation histories for four 
representative galaxy types. Because optical passbands provide poor leverage on 
star formation histories earlier than 1 Gyr prior to the epoch of observations, we have reduced the complexity of galaxy star formation 
histories to nonnegative combinations of discrete components such as those illustrated in these cartoons. For our purposes, each galaxy is 
modeled with six age-related components, as described in the text, with the oldest component starting at $z=5$ and continuing down to 
some time $t_{end}$ prior to the epoch of observation. Five younger, discrete components of duration 200 Myr allow for the broad possible 
range of complex histories in a galaxy's recent past. Each of these six components is quadrupled, with four different levels of extinction, 
$A_V\in \{0,0.5,1,2\}$ mag, leading to a total of 24 stellar components with non-negative contributions to the stellar mass of a galaxy. With 
redshift, metallicity, and $t_{end}$ as our gridded parameters, there are 24 stellar coefficients at each location in the grid. By abstracting 
star formation histories in this way, we have retained the essential information content of on-going star formation, intermediate-age 
populations, and an old, underlying stellar population. 
\label{fig:cartoon}}
\end{figure*}

\section{SED Fitting}
\label{sec:fitting}

With flux calibrated spectra and broadband photometry in hand, a suitable library of template spectra can now be employed to estimate 
spectrophotometric redshifts. In this section we describe the basis functions in our templates, implement a generalized maximum 
likelihood method and estimate confidence limits for the redshifts of each galaxy
along with several other parameters and properties.

Here we discuss the construction of the templates, using several continuum components, each derived from the \cite{maraston2005} models. 
The Kroupa (2001) initial mass function (IMF) was used, resulting in a median offset from ``diet'' Salpeter IMF (Bell \& de Jong 2001) of 0.04 
dex for SSP ages up to about 9 Gyr.

\subsection{Ingredients}
\label{subsec:ingredients}

Ideally, a set of spectral templates should not only span the broad range of redshifts a survey may probe, but also encompass the potential 
range of optical and near-IR properties of galaxies over those epochs. Satisfying this latter constraint cannot be done {\it a priori\/} 
as it is one of the chief goals of the project. However, models of evolving stellar populations have been used for modeling 
low-dispersion prism data for the purpose of recovering redshifts to $z\sim 1$ (see Patel et al. 2010). Our
method for CSI echoes that approach, and we describe it here.

Our templates are constructed as the superposition of several continuum components and multiple emission lines. The stellar population 
bases were derived from the \cite{maraston2005} models, using the Kroupa (2001) initial mass function (IMF)\footnote{This is equivalent 
to applying an offset of -0.04 dex to the stellar $M/L$ ratios of ``diet'' Salpeter IMFs (Bell \& de Jong 2001) for single stellar population 
(SPP) ages up to about 9 Gyr.}.  In building our templates we exploit the well known fact that for times $\tau \gappr 1$ Gyr after the 
cessation of star formation, optical flux no longer provides significant leverage on a galaxy's prior star formation history (Tinsley 1974). 
We adopt a simple set of stellar population basis functions: these components are sensitive to different timescales of a galaxy's star 
formation history; in combination, they reproduce the broad range of SED properties seen in normal galaxies.

The first base component is a constant star formation model with a starting epoch of $z_f=5$. The time at which star formation ceases for 
this component is a free parameter our analysis, a grid of values ranging from $10^9$ yr to $10^{10}$ yr prior to the redshift of observation 
(capped at $z=5$), whose values are spaced logarithmically with an interval of 0.10 dex. The metallicity of this base population is also 
gridded with values ranging from [Z/H]$=-1.2$ to [Z/H]$=0.6$, with an interval of 0.2 dex. Redshift is the final gridded parameter, ranging 
from $z=0.005$ to $z=2.0$ at intervals of $\Delta z=0.005$.  Superimposed on the base population are up to five piecewise constant SFR populations
that formed in 200\,Myr intervals centered on lookback times of $t=1,3,5,7,9\times 10^8$ yr,
with the same metallicity as that of the base. Four cartoon star 
formation histories built out of such components are schematized in Figure \ref{fig:cartoon}.  The SEDs of each of these stellar components 
are redshifted, convolved to the wavelength-dependent resolution of the instrument, and sampled at the wavelengths appropriate for the 
galaxy spectra.  Lastly, these are all replicated with extinctions of $A_V\in \{0.0,0.5,1.0,2.0\}$ mag in an effort to reproduce the effects of 
complexity in the geometries of dust attenuation than simple screen models.

In addition to the continuum components, we include several emission line components: (1) a single unresolved Gaussian emission line 
for [OII]3727\,\AA, (2) a blend of three Gaussians at [OIII]5007\,\AA, [OIII]4959\,\AA, and H$\beta4861$\,\AA, with ratios of $1:1/3:1/10$, to 
broadly mimic the typical line ratios seen in galaxies at the masses we expected (e.g. Kauffmann et al. 2003), (3) a Gaussian for H$\alpha6563$\,\AA, 
and (4) an unresolved emission line for MgI2799\,\AA.  Because of the increasing uncertainties in the flux calibration of LDP data beyond 8000\,\AA, 
no emission lines were allowed beyond that point in those fits. Given the higher quality of the UDP data in the far red, we extended this limit 
to 9000\,\AA\ in fitting UDP data, with the exception of [OII]3727\,\AA, which we accepted to 9400\AA.  These restrictions 
do not greatly affect the redshifts that are measured, but they do have some impact on the fit to the red continuua of low redshift galaxies. 
Note that the widths of the Gaussian profiles are fixed to that defined by the spectral resolution.

Thus there are in total 24 stellar continuum components and 4 emission line components, each redshifted, convolved to the 
wavelength-dependent resolution of the instrument, and sampled at the wavelengths appropriate for the galaxy spectra. The broadband 
magnitudes of each component are computed using the filter transmission and detector QE curves of the respective imaging instruments.

\subsection{Fitting for the Galaxy Components}
\label{subsec:fitting}

Before solving for the 28 coefficients at every location in the three-dimensional grid of redshift, metallicity, and termination time, we compute 
a simple constant rescaling of the IMACS spectrum to the broadband $griz$ photometry from the $D=4$ arcsec aperture. In general the 
colors of the IMACS spectra agreed within $\pm 10\%$ of the colors derived from the broadband photometry and adjusting for any
color offset did not significantly affect the measured redshifts.

Standard least-squares and $\chi^2$ minimization techniques are susceptible to bad data points and outliers, for example, from a few residual 
cosmic rays or unidentified bad pixels, especially given the small number of usable pixels in each spectrum ($\sim 150$). We opted to perform 
an iteratively reweighted non-negative least squares fit for the coefficients of the template fit at each location in the grid, using the weight 
function of Huber's M-estimator \citep{huber1981,zhang1997}. Doing so is equivalent to minimizing Huber's M-estimator itself,
given as $L_{Huber}$, with the associated weight function $W_{Huber}$:
\begin{eqnarray}
L_{Huber}(x) =
&\begin{cases}
x^2/2 & \text{if $|x| \le k$}\\
k(|x|-k/2) & \text{if $|x| \ge k$}
\end{cases}\\
W_{Huber}(x) =
&\begin{cases}
1 & \text{if $|x| \le k$}\\
k/x & \text{if $|x| \ge k$}
\end{cases}
\end{eqnarray}
where $k=1.345$. Near the optimum location, points retain the behavior of $L_2=x^2/2$ (i.e. least squares) while outside points retain the 
behavior of $L_1=|x|$ (i.e. minimum absolute deviation), leading to an estimator that is robust against outliers but sensitive to the details of the 
distribution close to the optimum location. In three iterations the routine effectively minimized this M-estimator,  instead of the $L2$ 
M-estimator, where $L2=\chi^2/2$. The result is a set of coefficients that are robust against any bad data present, as long as the fraction of bad 
pixels is $\lappr 25\%$.

With the rescaled IMACS spectra and the $ugrizJK_s$ photometry, we now solve for the template coefficients using iteratively 
reweighted non-negative least squares. The coefficients for each galaxy are stored, as well as several M-estimators for each grid location, for use 
in the final likelihood analysis (described below).

\subsection{Likelihood Analysis}
\label{subsec:likelihood}

Each location in the three-dimensional grid of redshift, metallicity, and the termination of star-formation for the base `constant-star-formation' 
model has a dozen coefficients, as well as goodness of fit metrics. With these grids of population coefficients and M-estimators, confidence
limits on all of the gridded parameters and derived properties, such as total masses, stellar masses, restframe colors, were
derived using a likelihood function
\begin{equation}
\begin{split}
d\mathcal{L} = &dz d\log t_{end} d[Z/H] P(z,\log t_{end},[Z/H])\times\\
&\exp\bigl[{-L_{Huber}(z,\log t_{end},[Z/H])}\bigr]
\end{split}
\end{equation}
where $P$ is the defined set of priors at a given redshift, termination time, and metallicity. In our likelihood analysis we found that using
Huber's M-estimator gave results that were most robust to bad pixels without biasing the contributions of emission lines, though others, for
example, ``least-power'' or ``Fair'' \citep[see, e.g.,][]{zhang1997}, were also very effective. We caution against using M-estimators that 
are insufficiently concave, as pixels ``contaminated'' by real (but narrow) features such as unresolved emission lines are attributed less weight 
than continuum pixels (by definition). With the three dimensional likelihood functions, and 24 SFH + 4 emission line parameters at each location 
in the grid, we have computed confidence limits for every parameter, and for several derived combinations of parameters such as stellar mass 
or a broad range of restframe colors. With these confidence limits, and a broad array of data quality measures, such as $S/N$ ratios, minimum 
$\chi^2$, $\chi^2$ at the peak of the likelihood function, and the confidence limits themselves, we can identify the highest quality sample possible 
with the current dataset.

\begin{figure*}[htbp]
\centerline{
\includegraphics[width=6.0in]{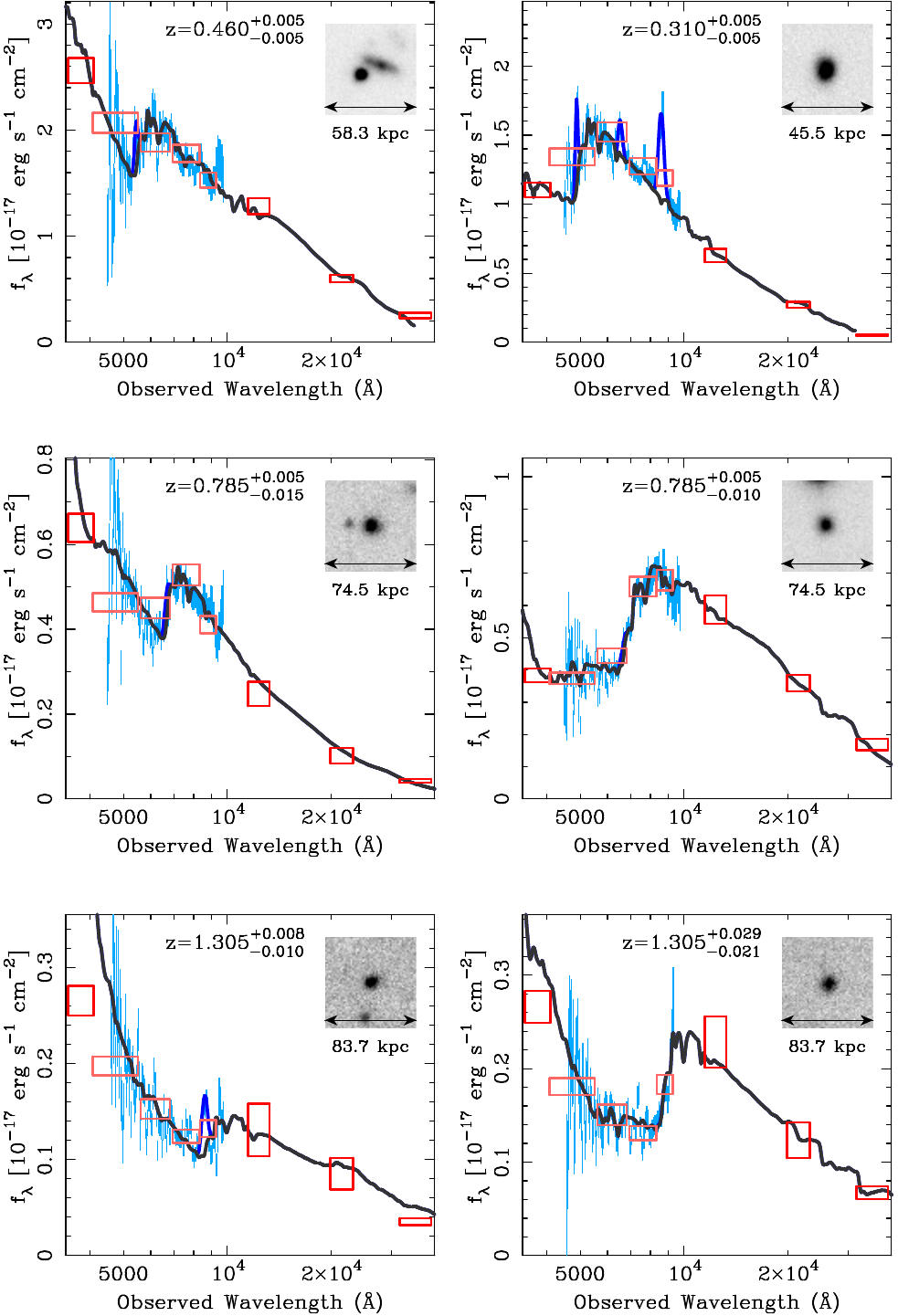}}
\caption{Example SEDs for blue galaxies, along with their CSI redshifts and 95\% confidence limits.
The IMACS spectra are shown in cyan. The red boxes mark the broadband flux measurements in
$ugrizJK_s[3.6]$, noting that the SED fitting was performed using $ugrizJK_s$. The best-fit stellar
population models are shown with thick black lines. When the data require emission line components
in the fit, these are shown using dark blue. Inset we show the CFHTLS $z$-band images of the
galaxies. In particular, we note the top-left and bottom-left, two galaxies for which broadband
photometry alone would be insufficient to provide redshift estimates. Low dispersion spectroscopy is
key to isolating Balmer breaks and emission lines in blue galaxies, leading to redshift
uncertainties with very little dependence on spectral type.
\label{fig:examples1}}
\end{figure*}

\begin{figure*}[htbp]
\centerline{
\includegraphics[width=6.0in]{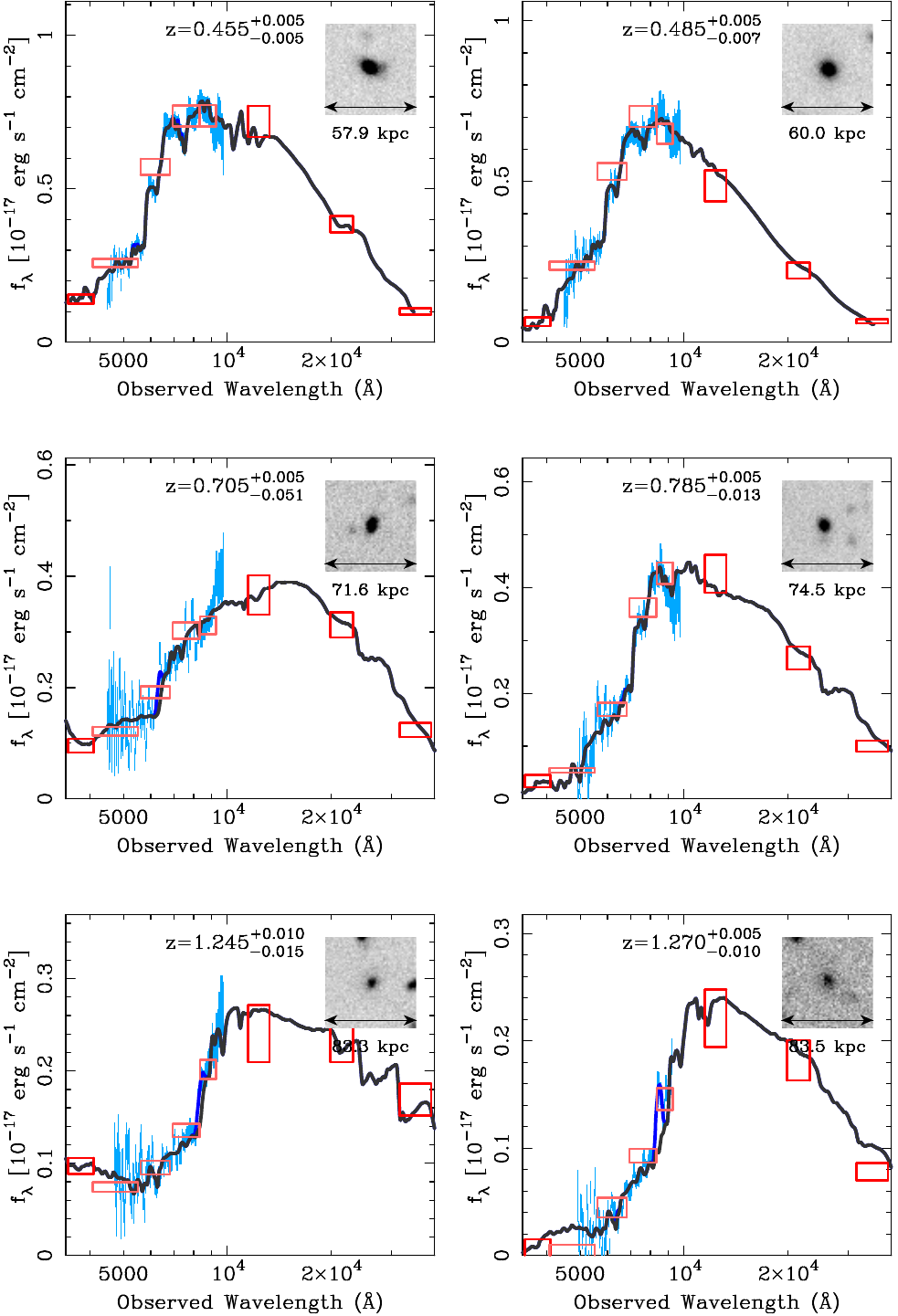}}
\caption{Same as in Fig. \ref{fig:examples1} but for red galaxies spanning a range of redshifts in CSI.
Note the structure in the SEDs traced by both the data and the models outside of the 4000\AA\ break.
\label{fig:examples2}}
\end{figure*}


\subsection{Defining the High Quality Sample}
\label{subsec:define}

Several criteria are used to eliminate substandard results. A number of measures were judged to be helpful, including $\chi^2$ for the best-fit 
template at a given $S/N$ ratio, the number of pixels not flagged as bad or contaminated, or unphysical inferred rest frame colors. The final 
criterion used to define our high quality sample was to restrict the set to those galaxies with formal 95\% uncertainties in rest frame 
$M_g$ of $\Delta_{95}M_g < 2.0$ mag (equivalent to $\sigma_{M_g}< 0.5$ mag). The result is a relatively clean sample of 37,000 galaxies, for which we have 44,000 spectra, over a field 
of 5.3 degs$^2$. For every spectrum, we have marginalized over redshift, stellar mass, rest frame colors, stellar population parameters as
per the components in the SED fitting, and emission line luminosities, estimating 68\%, 90\% and 95\% confidence intervals using the likelihood 
functions.

Example results of the SED fitting illustrating a broad range of redshifts and spectral types
from our high quality sample are shown in Figures \ref{fig:examples1}
and \ref{fig:examples2}.

Given that our sample is more effectively reaching higher redshifts at a given stellar mass
than optically selected surveys such as DEEP2 and PRIMUS, we now turn to
characterizing the accuracy of our redshifts.

\subsection{Success Rates and Completeness}

\begin{figure*}[t]
\centerline{
\includegraphics[width=6.5in]{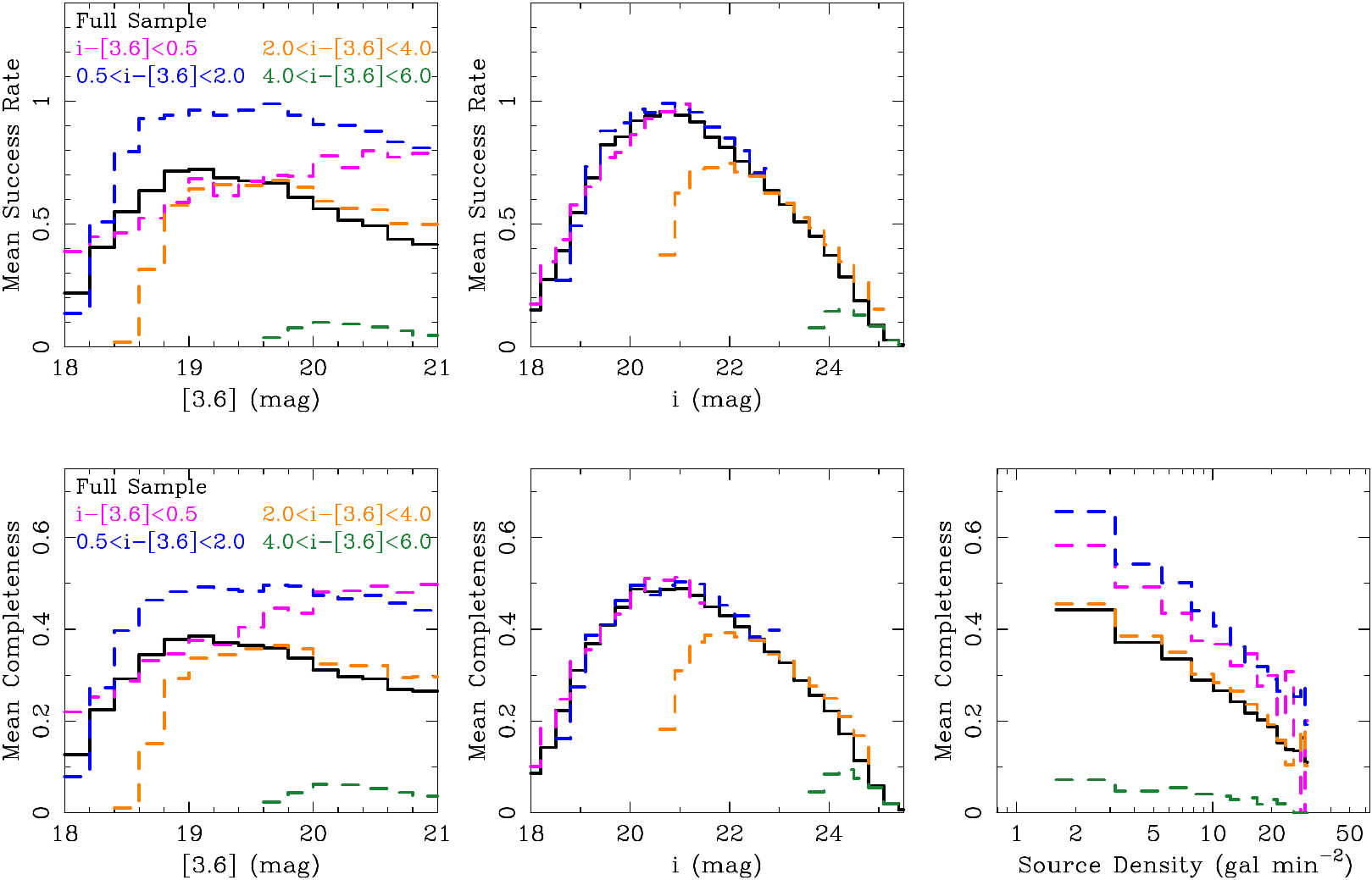}
}
\caption{(top) Average success rate
as functions of $3.6\mu$m and $i$ magnitude, where success rate refers to the fraction of slitlets cut that
led to a high quality redshift measurement. (bottom) The average completeness
as functions of $3.6\mu$m and $i$ magnitude, where completeness is the fraction of galaxies
above the $3.6\mu$m flux limit within the 5.3 degs$^2$ area that have a high quality CSI
redshift. At a given $3.6\mu$m flux, the completeness is a
fairly strong function
of color, such that it very nearly depends on $i$-band magnitude alone. We also plot the mean
completeness as a function of local
source density computed on scales of $2/3$ arcmin in an effort to account the portion of our
incompleteness due to slit collisions.
\label{fig:comp}}
\end{figure*}

In a spectroscopic galaxy survey, the degree to which the population is sampled can be characterized
by two quantities: the success rate (fraction of \emph{targeted} galaxies for which usable spectra are obtained) and the
completeness (fraction of the original \emph{photometric} that is observed and usable spectra obtained).
The latter quantity necessarily is necessary dependent on the former, and includes the effect of slit collisions.
In Figures \ref{fig:comp} (top) and (bottom) we plot success rates and
completeness as functions of $3.6\mu$m, $i$, and $r$ magnitudes. For
galaxies brighter than $r=25.5$ mag the current pipeline produces a mean success rate of 70\% and
mean completeness of 38\%. In detail we model the completeness as a 2D function of both
$3.6\mu$m magnitude and $i-[3.6]$ color, and for illustrative purposes we have overplotted the mean success rate and
completeness within broad color bins (though our subsequent analysis employed finer color bins
when modeling the incompleteness).

Spectroscopic surveys such as CSI that intend to study the density dependence of galaxy properties must carefully model the effect of 
slit collisions on the completeness. To accomplish this we divided our sample into bins based on local source density.  The footprint of a 
CSI spectrum is $\sim 200$ pixels long, or $2/3$ arcmin. For every target and observed galaxy, we compute the local source density 
within $4/9$ arcmin$^2$. In Figure \ref{fig:comp} we show the mean completeness as a function of source density computed on this 
scale. This monotonic function of density is taken to represent the effects of slit collisions.
We have also computed the completeness 
functions as coarse trivariate functions of source density, $[3.6]$ magnitude, and $i-[3.6]$ color in
and found no statistically significant variation in the
bivariate ($[3.6]$, $i-[3.6]$) 
completeness function with source density, other than the normalization, as shown in 
Figure \ref{fig:comp}.

Using the trend of the mean completeness as a function of source density, and the mapping of the completeness as a bivariate function of 
$z$-band magnitude and $z-[3.6]$, we can directly multiply the two for every galaxy in the final sample. In subsequent, and relevant analysis,
one can employ weights defined as the inverse of these completeness estimates. For those galaxies with multiple observations, we currently 
divide these weights by $N_{obs}$, the number of observations, though more complicated procedures may be devised at a future date.
We have tested the validity of these density-dependent completeness corrections by recomputing the local source densities based on the 
high-quality spectroscopic sample alone. On the scales of $2/3$ arcmin, the agreement is very good, with a scatter of approximately 
$0.3$ dex. On larger scales, of 2 or 4 arcmin, the agreement is even better with scatters of $<0.1$
dex at a given density.
These corrections, and the accuracy with which they can reproduce the original source densities, are crucial for ensuring 
confidence in the galaxy group mass estimates derived below.

\subsection{Redshift Uncertainties}
\label{subsec:accuracy}

We have
three different methods at our disposal for evaluating the accuracy of CSI redshifts:  (1) directly
comparing our results with several hundred optically-selected high-resolution spectroscopic redshifts in the VVDS sample \cite{lefevre2003};
(2) deriving uncertainties using the large number of objects with repeat measurements;
and (3) employing the \cite{quadri2010} pairwise velocity approach, in which galaxy groups and
large scale structures mean that large numbers of galaxies lie at the same redshifts, effectively providing repeat redshift measurements of cosmic structures.
All three methods give consistent estimates of the redshift uncertainty to $r\sim 25$ mag and/or $z=1.3$. At fainter magnitudes and
higher redshift there is insufficient overlap to empirically characterize the redshift errors using
the VVDS. However, to these limits, these checks on our redshift uncertainties all show good
agreement.

\subsubsection{Comparison with VVDS}

\begin{figure*}[t]
\centerline{
\includegraphics[width=5.5in]{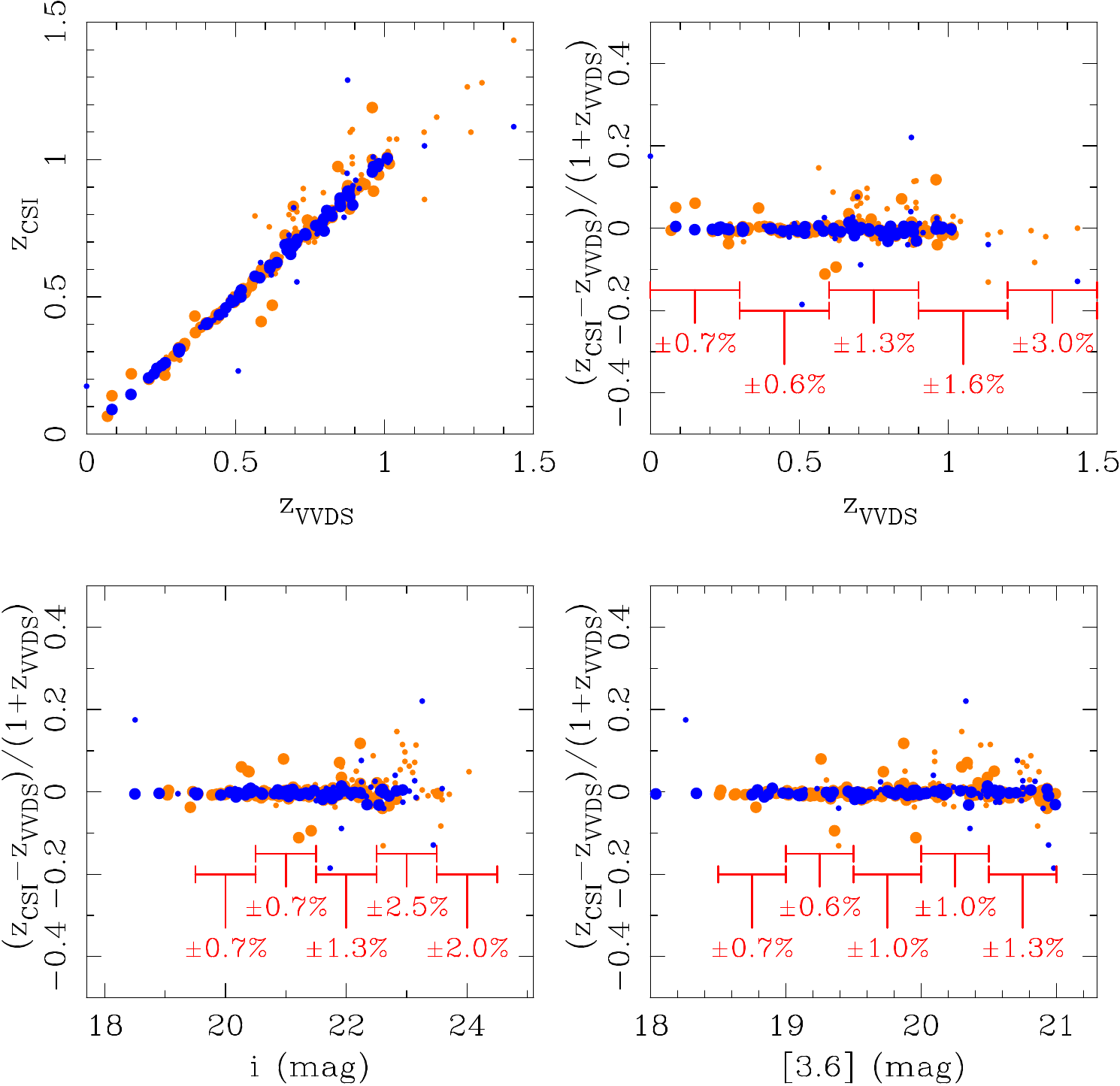}
}
\caption{(top left) CSI redshifts plotted against VVDS redshifts with quality flags 3,4,5. Larger
symbols indicate higher quality VVDS redshift flags. In orange and blue we show those galaxies
observed with the LDP and UDP respectively. (top right) Fraction redshift error vs. VVDS redshift.
Numbers shown with bins along the x-axis show the $1\sigma$ estimate of the scatter, computed using
$\sigma=1.48\times$ MAD \cite[median absolute deviation;][]{beers1990}.
(bottom left) and (bottom right) Fractional redshift error vs. $i$-band, and $3.6\mu$m
magnitude. For galaxies with high quality VVDS redshift flags, we find a $1 \sigma$ scatter of 1\%
in $\delta z/(1+z)$, and that for galaxies between $0.7<z<1.0$, 94\% have fractional redshift errors
$\delta z/(1+z)<5\%$. 
\label{fig:vvds}}
\end{figure*}

In  our sample of galaxies with high quality redshifts, there are 400 with VVDS spectroscopic measurements; 
Figure \ref{fig:vvds} compares these redshifts.  We have visually inspected and verified redshifts of the VIMOS spectra of these 
objects common to the VVDS and CSI, in a few cases revising VVDS redshifts and quality flags. For VVDS objects flagged as 
`low quality' (3 \& 2), we were not able to verify all redshifts; we nevertheless include them here, using smaller points in 
Figure \ref{fig:vvds}.

For the 191 galaxies with high quality VVDS measurements, 91\% of the redshifts agree to within $\delta z/(1+z)<0.02$, with 96\% within 
$\delta z/(1+z)<0.05$. The standard deviation for this subsample is $\sigma_z/(1+z)=0.008$
\cite[using $\sigma=1.48\times$ MAD median absolute deviation;][]{beers1990}.
For the 138 with poor quality flags in the VVDS subsample
the scatter is $\sigma_z/(1+z)=0.014$, with 86\% agreeing to within $\delta z/(1+z)< 0.05$. In Figure \ref{fig:vvds}
we show the $1 \sigma$ scaters in different magnitude bins also derived using $\sigma=1.48\times$ MAD.
Note that these values include a number of galaxies with low quality VVDS redshifts. For galaxies down to the PRIMUS
selection limit of $i=23$ mag, the scatter is $\sigma_z/(1+z)=0.009$. Fainter than $i=23$ mag,
the scatter rises to $\sigma_z/(1+z)=0.025$. For galaxies selected by CSI to be between $1<z<1.5$, we find
$\sigma_z/(1+z)=0.033$, though the distribution is non-Gaussian and the subsample largely contains objects with
low quality VVDS redshifts.

The comparisons between our redshifts and those made available by the VVDS have allowed us to confirm that our procedures 
are working well. However, a larger numbers of galaxies is required to accurately characterize redshift uncertainties as a function of 
magnitude, color, or spectral type.
Fortunately, we have two additional tests to help us probe the accuracy of our redshifts.

\begin{figure*}[t]
\centerline{
\includegraphics[width=6.0in]{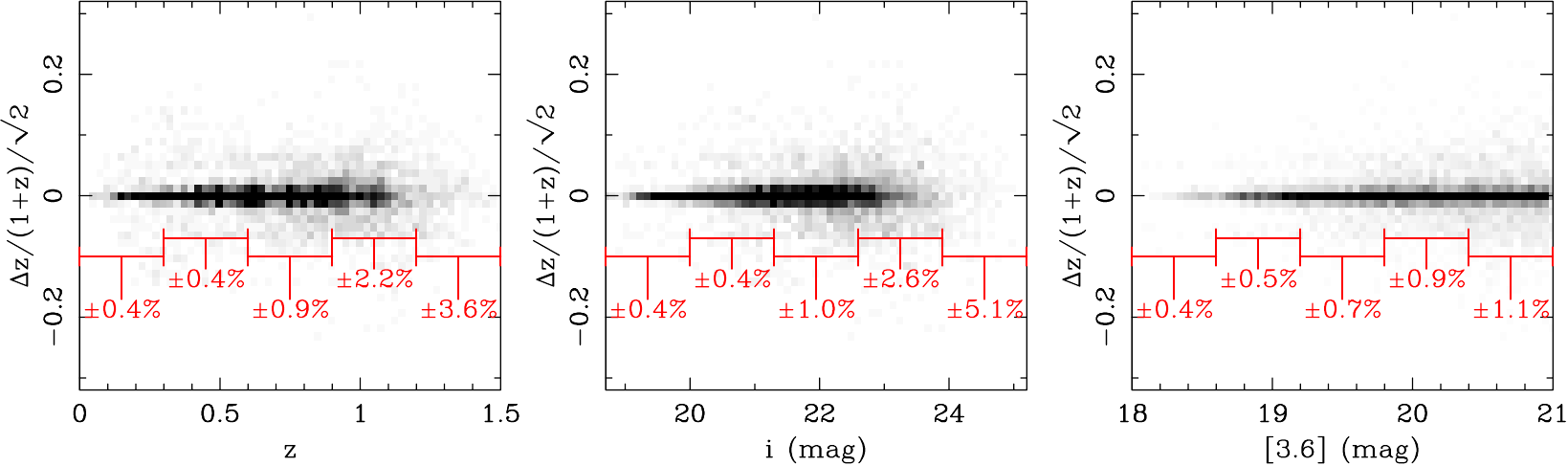}
}
\caption{Fractional redshift differences between repeat measurements, 
scaled by $\sqrt{2}$, as functions of redshift, $3.6\mu$m magnitude, and 
$i$-band magnitude.
For example, at $0.6<z<0.9$, the half-width of the 68\% confidence interval 
is $\sigma=0.9\%$. At $0.9<z<1.2$ we find  $\sigma=2.3\%$. As can be seen 
in the third panel, the width of the distribution is a stronger function of optical 
magnitude than near-IR magnitude, because our SED fitting is dominated by 
the IMACS spectroscopy.  We find $\sigma=2.6\%$ at 22.5 mag $<i<$ 24 mag, 
and $\sigma=5\%$ at 23.5 mag $< i <$ 24.8 mag. 
\label{fig:repeats_z}}
\end{figure*}

\subsubsection{Uncertainties Derived from Repeat Measurements}

The data from individual slit masks were reduced independently, and so repeat measurements can be compared. These comprise
approximately 20\% of the measured redshifts, providing a sample large enough to be analyzed as a function of galaxy properties, 
such as magnitude, color, stellar mass, or redshift. Without prior constraints or measurements for the derived properties of these
objects we adopt the mean of multiple measurements when plotting or analyzing the distributions as functions of redshift. In 
Figure \ref{fig:repeats_z} we plot the 68\% fractional redshift differences against redshift, as well as $3.6\mu$m and 
$i$-band magnitudes. As a function of redshift the half-width of the 68\% confidence is $0.9\%$ at $0.6<z<0.9$, $2.2\%$ for $0.9<z<1.2$. 
As can be seen in the third panel, the width of the distribution is a stronger function of optical magnitude than near-IR magnitude because 
our SED fitting is dominated by the IMACS spectroscopy. Currently, we find $\sigma=3\%$ at 22.5 mag $<i<$ 24 mag, and $\sigma=5\%$ 
at 24 mag $< i <$ 25.2 mag. Planned refinements in the pipeline are expected to improve the fidelity of the extracted spectra for
such faint sources as the CSI Survey progresses. 

\begin{figure*}[t]
\centerline{
\includegraphics[width=7.0in]{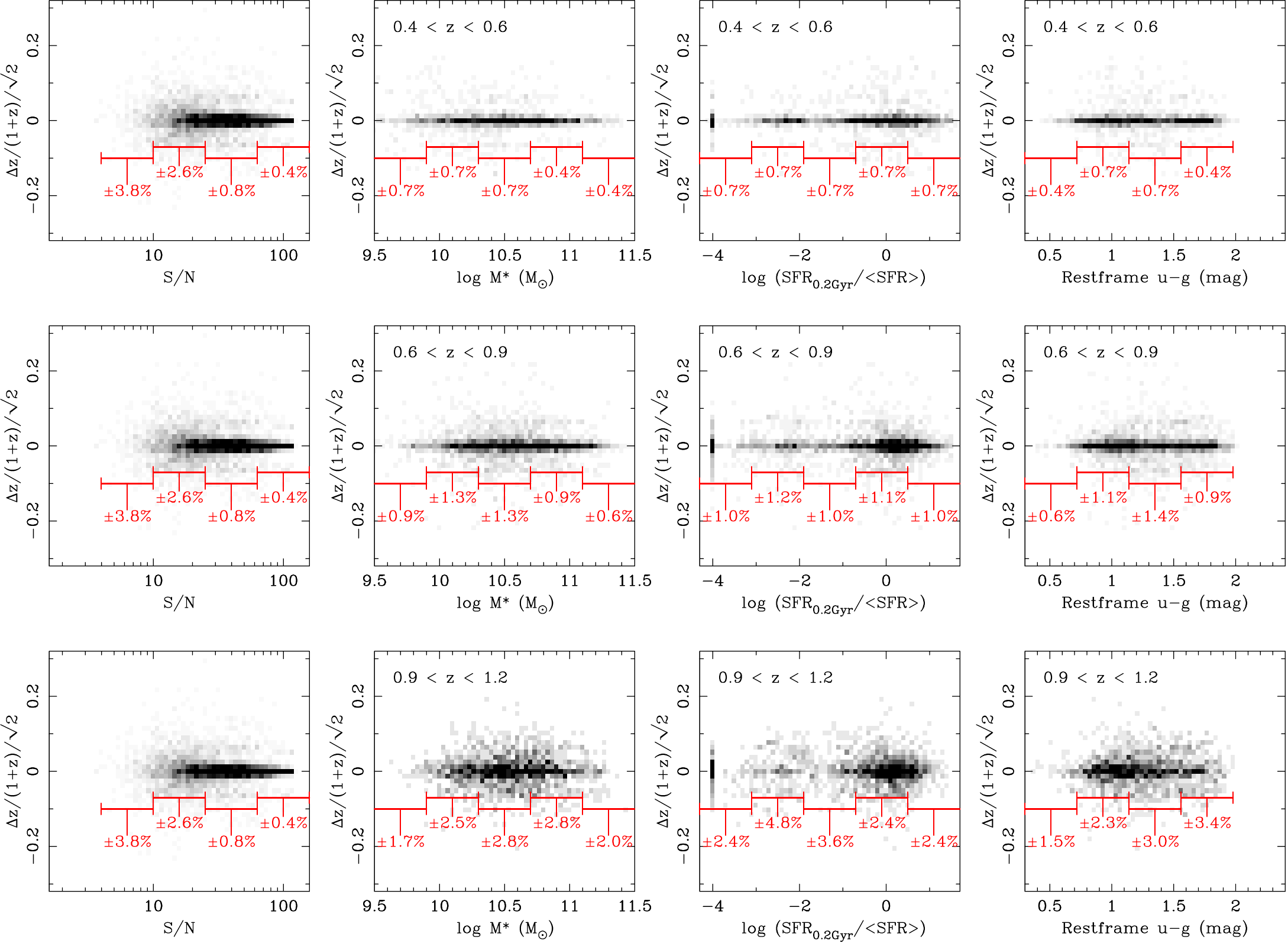}
}
\caption{Fractional redshift differences between repeat measurements, 
scaled by $\sqrt{2}$, as functions of $S/N$ ratio (75th percentile of the prism spectra), stellar mass, relative 
star formation activity, and restframe $u-g$ color. There is very little dependence of 
the redshift errors on spectral type or color, unlike what is typically seen 
for photometric redshifts.
\label{fig:repeats_z2}}
\end{figure*}

The redshift uncertainties are dissected further in Figure \ref{fig:repeats_z2} by breaking up the
sample into three redshift ranges and plotting against $S/N$ ratios of the IMACS spectra, stellar
mass, star formation activity, and rest frame $u-g$ color. From these panels one can see that the
redshift errors are not strongly correlated with spectral type or color, unlike the typical
dependencies seen in photometric redshifts.
There is a rather small, subtle dependence on spectral
type: quiescent galaxies (those with a negligible specific star formation rate and hence strong Balmer/4000\AA\ breaks) and strongly
starforming galaxies have the greatest precision, while galaxies with modest or intermediate SFR
have degraded redshift precision.
Weak [OII]3727\,\AA\ emission at high redshift and observed at low resolution can appear blended with the
4000\AA\ break break and produce spectra that are reasonably well fit by an SED with a Balmer break
and no [OII]3727\,\AA. More specifically, [OII]3727\,\AA\ is imperfectly correlated with the SFR
inferred from the stellar continuum, leading to an additional uncertainty in the redshifts at the
$\sim 1-2\%$ level, depending on the $S/N$ ratio of the spectra, and even when not particularly
statistically significant.
This effect is substantially less pronounced for UDP data, where the higher resolution of the  disperser shifts the 
wavelength of such [OII]3727\AA\ confusion farther to the red. This power of the higher resolution
UDP is illustrated in the last SED of Figure \ref{fig:examples1}, in an old galaxy at $z=1.27$ with
[OII]3727\AA\ line emission \citep[likely a LINER, e.g.][]{lemaux2010}.
Again, this problem with low resolution spectroscopy is only important for ``green''
galaxies;  the bulk of the galaxy population, lying in the blue cloud and red sequence 
(with strong [OII]3727\,\AA\ and Balmer breaks respectively), is unaffected.

While the abundance of repeat spectroscopic measurements has allowed us to characterize the redshift
uncertainties, repeated spectra of any given object are fine-tuned with the same broadband photometry as described in Section \ref{subsec:fitting}.
For some galaxies this can in principle introduce artificially tight correlations between the repeat spectroscopic measurements.
Presumably, this issue becomes a serious problem only if the broad-band photometry is allowed to have greater weight in the
likelihood analysis. This never occurs in our procedures, in particular because the number of data points in the UDP and LDP
spectra far exceed the number of photometric bands. Even when the $S/N$ ratios of the spectra are low, binning these data over
the photometric bandpasses yield $S/N$ ratios that still exceed what obtains from the photometry for those fainter galaxies.
Even with these reasons to
trust the analysis of the repeat measurements, we employ yet another means to assess the
redshift errors in the next section.

\subsubsection{Uncertainties from Pairwise Velocity Distributions}

Because galaxies largely exist in close pairs, in groups, in clusters, and in even larger coherent structures, all of which have velocity 
widths smaller than the redshift errors we expect in our data and analysis, the distribution of redshift errors can be inferred statistically 
from the distribution of pairwise redshift differences in a given sample. This technique was first explored and described by
\cite{quadri2010} in an effort to derive the redshift uncertainties in photometric redshift surveys, where a lack of repeat measurements 
and limited templates can hamper one's ability to estimate uncertainties. Any methodology relying explicitly on clustering requires 
dense spatial sampling in order to extract a high $S/N$ signal from the pairwise velocity histograms.  While photometric redshift surveys, 
by definition, result in the highest possible  spatial sampling, the spatial sampling of CSI is limited by slit collisions.  This limits
the number of galaxy pairs available to perform this analysis, compared to what would be obtained in a purely photometric 
redshift survey.

We divided our sample into subsets based on $r$-band magnitude and redshift and constructed pairwise velocity histograms. For the 
subsamples based on magnitude, we used those galaxies brighter than a given bin as a reference set, assuming those galaxies have 
more precise redshifts. While \cite{quadri2010} chose to randomize the positions in their dataset to correct for close pairs that arise only in projection, we opted to randomize the 
galaxy velocities since our slit-placement constraints lead to nontrivial spatial dependencies.
Care was taken to properly normalize the histograms derived from the randomized distributions, as the number
of galaxies in them, by
definition, is too large by the number of galaxies within coherent velocity structures.
Because this number cannot be precisely 
defined {\it a priori\/}, it must be deduced from the data --- from the wings of the histograms. We made several realizations 
to reduce the Poisson noise imposed by the randomization process.

\begin{figure}
\centerline{
\includegraphics[width=2.0in]{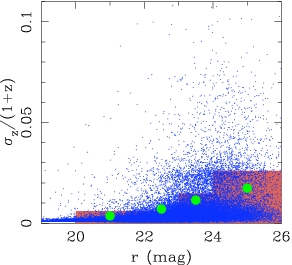}
}
\caption{A comparison of the redshift uncertainties derived from the likelihood functions with the
mean uncertainty as a function of magnitude derived from the pairwise velocity histogram method of
\cite{quadri2010}. In blue we plot  $1 \sigma$ redshift uncertainties for the individual galaxies.
In green we plot the expected standard deviation of the
pairwise velocity histograms from the blue points. The red shaded bars trace the measured standard
deviations from the pairwise velocity histograms as functions of
magnitude. The comparison is quite good, suggesting that the uncertainties from our redshift fitting procedures
are robust.
\label{fig:quadricompare}}
\end{figure}

The results of these experiments are shown in Figure \ref{fig:quadricompare} as the red histogram.
Overplotted are the $1\sigma$ confidence limits on our redshifts, shown in blue, and computed by
dividing the  $95\%$ confidence limits by four. The green filled circles are the expected standard
deviations of the pairwise velocity histograms for galaxies with redshift uncertainties shown in
blue.
Overall the \cite{quadri2010} method has reinforced the validity of the
redshift uncertainties derived from our generalized likelihood analysis down to faint magnitudes, and from our comparisons to the relatively bright VVDS sample.

This result should be of general interest, since the \cite{quadri2010} method was in part devised to remedy the deficiency that 
photometric redshift surveys do not provide repeat measurements of a given galaxy.  However, any spatially dense survey will 
provide repeat measurements of (relatively) cold cosmic structures, so the \cite{quadri2010} methodology remains a
valuable technique for any spatially dense spectroscopic or photometric redshift survey --- any survey
where one does not have repeat observations with which to characterize uncertainties empirically.

Based on these three methods, we conclude that our random redshift uncertainties have been reliably characterized,
and also that systematic uncertainties appear to be minimal. In
particular we stress that our redshift errors are not strongly dependent on
relative star formation activity, spectral type, or color. As discussed
by \cite{quadri2012}, redshift errors that are strongly correlated with
spectral type will bias estimates of local galaxy density if care is not taken to ensure that velocity
windows or velocity linking lengths can encompass both red and blue galaxies with equal probability.
Given CSI's primary goal of characterizing galaxy properties as a function of environment, the relative insensitivity of
our redshift errors on spectral type is a crucial point.

\begin{figure}[htb]
\centerline{
\includegraphics[width=2.5in]{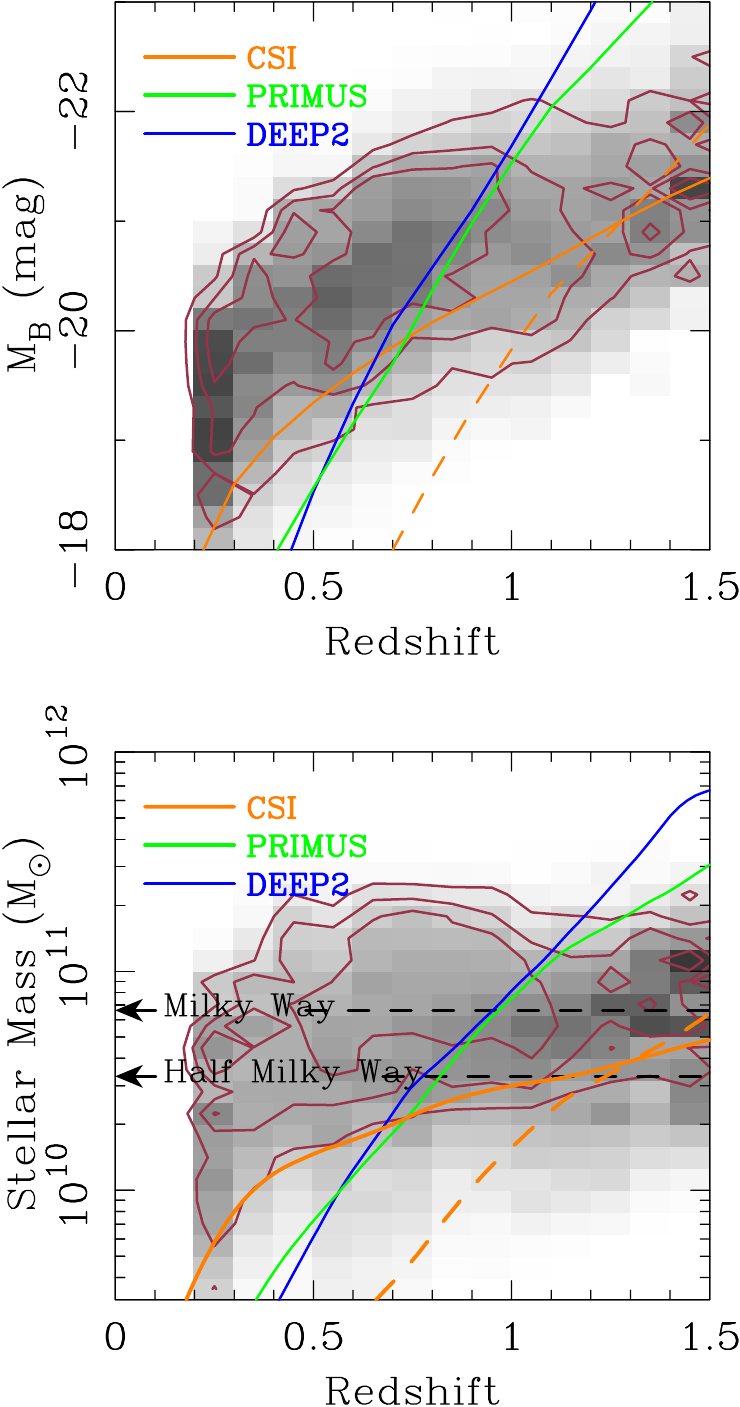}
}
\caption{The distribution of our $3.6\mu$m-selected galaxies in (top) restframe
$B$-band magnitude and (bottom) stellar mass, as functions of redshift, where
we have normalized the distribution at each redshift by the number counts in
each redshift bin.
The green and blue lines indicate the approximate loci for old stellar
populations at the limits of the PRIMUS and DEEP2 surveys. In orange, we
show the CSI selection limit of $3.6\mu$m $=21$ mag (solid) and the effective
optical limit of $i=24.7$ mag (dashed).
The maroon contours trace the distribution of ``quiescent'' galaxies, as
selected by their restframe $(U-V)$ and $(V-J)$ colors \citep{williams2009}.
Selection in the near-IR has ensured that CSI does not {\it a priori\/} exclude the
bulk of passively evolving systems at $z\simgt 1$. Meanwhile, the observing strategy,
reductions, and first analysis of the data enable CSI to reach more than a magnitude
fainter in the optical than DEEP2 or PRIMUS.
\label{fig:bz}}
\end{figure}

\begin{figure}[htb]
\centerline{
\includegraphics[width=2.5in]{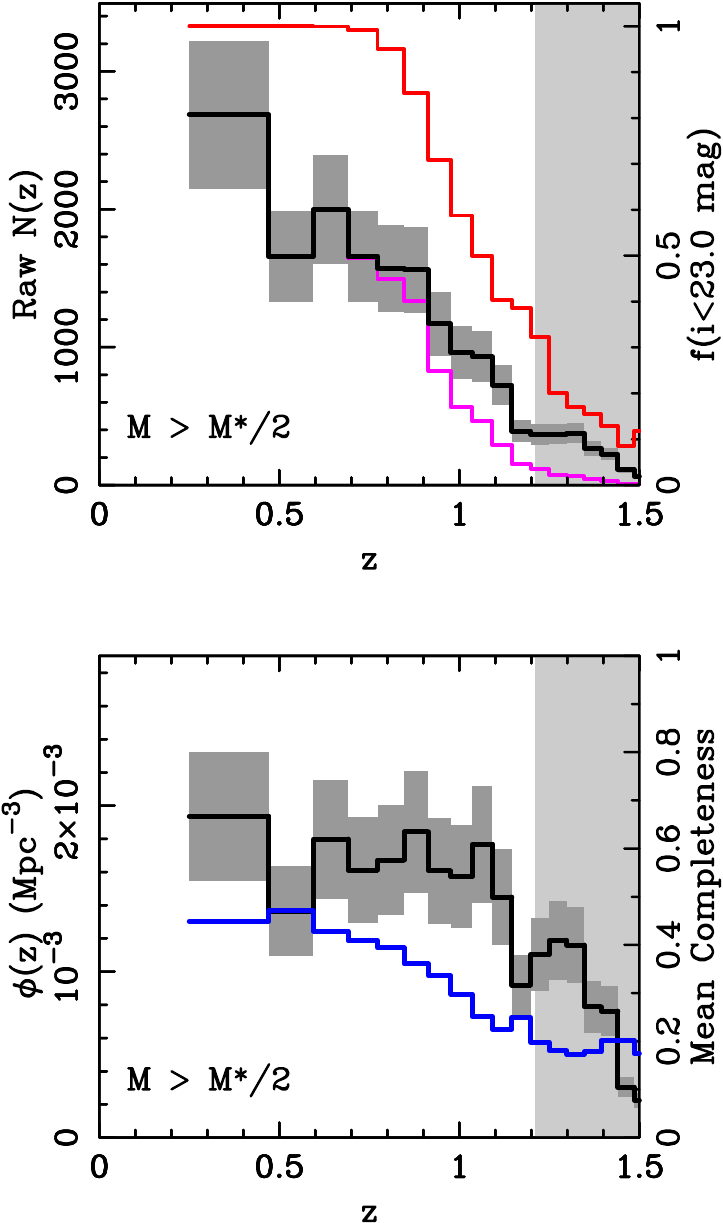}
}
\caption{
(top) Distributions of high quality redshifts for galaxies more massive than half that of the Milky
Way. The redshift bins are defined by a constant comoving volume in which cosmic variance is $\sim
20\%$ for red galaxies \citep[dark gray boxes;][]{somerville2004}. The apparent decline in $N(z)$
above $z=0.7$ is due to an increase in spectroscopic incompleteness (i.e. a declining success rate,
see the bottom panel). (bottom) Applying completeness corrections, we convert $N(z)$ to the comoving
number density of galaxies as a function of redshift. We plot the mean completeness as a function of
redshift in blue. The light gray shaded region beyond $z=1.2$ in the right-hand panel shows
approximately where the faint optical magnitude limit of the current analysis cuts the sample on the
red sequence (see Figure \ref{fig:masslimits}).
\label{fig:nz}}
\end{figure}

\section{Discussion}
\label{sec:general}

We conclude our discussion of the survey
by presenting a broad overview of the CSI sample, by tracing basic properties of these
Spitzer-selected galaxies over time. We specifically highlight the selection limits in order to give
the reader a better sense 
of the depth of the sample in redshift, luminosity, and stellar mass.

Figure \ref{fig:bz} shows the distribution of (top) rest frame $B$-band magnitude and (bottom) stellar mass as functions of redshift.
The dark red contours specifically outline the distribution of quiescent galaxies, as selected by their restframe $U-V$ and
$V-J$ colors \citep{williams2009}.
The CSI $3.6\mu$m flux limit is shown by the solid orange lines, along with
our effective optical limit of
$i\approx 24.7$ mag as dashed orange lines. This $i$-band limit, where our spectroscopic completeness falls below 1/3 of the
maximum (of 50\% at $i=21$ mag, see Fig. \ref{fig:comp}),
dominates over the IRAC selection
at $z>1.2$. In these plots, the sample becomes irreparably incomplete for galaxies at masses or magnitudes below the lines.
These figures show that
the goal of producing a large, cosmologically representative set of galaxies back to $z\approx 1.5$
is being achieved. CSI is probing galaxies with the
present day stellar mass of the Milky Way to a time when the universe was 9 Gyr younger than today
and without explicitly excluding galaxies with low $M/L$ ratios.

For comparison, we plot
the DEEP2 and PRIMUS magnitude limits of $r=24.1$ mag and $i=23.0$ mag using the blue and
green lines, respectively. These surveys can probe unbiased samples down to the stellar mass of the
Milky Way back to $z=0.9$, and half the mass of the Milky Way to $z=0.75$. CSI's additional reach
to $z=1$ for galaxies with half the mass of the Milky Way may only reach 1 Gyr farther back in
cosmic time, but the comoving volume covered within a constant area on the sky is nearly double that
of PRIMUS.
Such gains in sensitivity highlight the importance of an infrared 
selection in crafting a minimally unbiased picture of galaxies since $z=1.5$.

Figure \ref{fig:nz} (top) shows the distribution of high quality redshifts for galaxies with stellar masses $M>M^*/2$,
using redshift bins estimated to have constant cosmic variance of $\sim 20\%$
\citep[shown in dark gray]{somerville2004}. A large fraction of the decline in $N(z)$ is due to
our declining success rates and completeness at fainter optical magnitudes (as shown by the blue
lines in the bottom panel). As can be seen Figure \ref{fig:bz}, however, selection at brighter optical
limits, however, would have dramatically exacerbated the incompleteness.
Cutting at $i=23$ mag (the PRIMUS magnitude limit) would have eliminated $\sim 50\%$ of
those galaxies at $z=1$ in CSI. Selection at those wavelengths that more closely trace stellar
mass, and a methodology for measuring redshifts that does not depend crucially on the presence of
emission lines at faint magnitudes, ensures, at least {\it a priori\/}, that the resulting samples
are free from bias against galaxies with low $M/L$ ratios at fixed mass.

We have applied completeness corrections and
converted our $N(z)$ figures to estimates of the comoving number densities of galaxies as
functions of redshift in this stellar mass range and plot them in the bottom panels.
The dark gray, again, shows the estimated cosmic variance of $\sim 20\%$, which dominates the
uncertainties ($\sqrt{N}/N$ is small).
The light gray shaded region shown at high redshift in the bottom right indicate redshifts where our
faint magnitude limits impinge on the low mass end of this subsample and the selection biases become important. The mean of
our completeness estimates as functions of redshift are plotted in blue.

Detailed analysis of the evolution in number densities of galaxies over cosmic time is beyond the
scope of this first paper, which has primarily focused on the motivation behind the survey and discussion
of the handling of the data.
However, Figure \ref{fig:nz} reinforces what previous deep surveys covering smaller fields-of-view have
found:  fairly modest evolution in the number densities of normal galaxies since $z=1$
\citep[e.g.]{drory2009,marchesini2009,ilbert2010}.
In subsequent publications from
the Carnegie-Spitzer-IMACS Survey, we will begin dissecting the dataset in order to
identify galaxy groups and trace the evolution of the mass function of such groups since
$z=1$ \citep{williams2012}, and begin characterizing the star formation activity and histories of
galaxies in groups of different mass scales as a function of time.

\section{Summary}
\label{sec:summary}

We have described the methodology and initial data of an ambitious new survey of approximately $2\times 10^5$ galaxies 
over a volume that encompasses the last 9 Gyr of cosmic history.  By selecting a sample at 3.6$\mu$m from
Spitzer-IRAC, the CSI survey is the most uniform to-date in terms of limiting mass as a function of redshift, and
will ultimately cover an unbiased volume comparable to SDSS but at $0.5<z<1.5$.  The power of
the \emph{IMACS} spectrograph with low-resolution prisms allows us to survey large volumes efficiently and with sufficient
spectral resolution to detect large-scale structure and to measure emission lines from strongly star-forming galaxies and AGN.  CSI provides comparable redshift accuracies for red sequence and blue cloud galaxies; a significant advantage
compared to many broadband photometric---and even some spectroscopic---studies.  By combining our low-resolution 
spectrophotometry with extended broad-band photometry and sophisticated SED modeling, CSI bridges the
gap between surveys that are deep enough to probe galaxies below M$^*$ but are small in sample size and volume, and
those large-sample, large-volume surveys that do not reach typical, Milky Way-like galaxies that are the main event in
the history of cosmic evolution. In forthcoming papers we will update with more detail what we have
learned from the first batch of CSI survey data.

Once the three survey fields of CSI have been completed at the end of 2012,
we will have probed a volume equivalent to the SDSS, enabling a vast number of studies of
distant galaxies. One of the highest priority goals will be to take advantage of this volume to simultaneously measure
the evolution of the group mass function \citep[e.g.][]{williams2012},
and the dependence of star formation on environment \citep[e.g.][]{kelson2012}.
By doing so CSI will uncover the extent to which the growth in the number density of passive galaxies
over the past $2/3$ of the lifetime of the universe can be attributed to group-related processes such as, for example, galaxy-galaxy
interactions (leading to differential number density evolution between groups and galaxies),
and/or a decrease in the available fuel for on-going star formation (leaving galaxy number densities
unchanged). Given that most galaxies in groups are ``red and dead'' even at $z\sim 1$ 
\citep{kelson2012}, the mass selection, high spectral coverage, and color-insensitive
redshift accuracy of CSI will all be critical elements in characterizing relatively poor groups at
high redshift. Subsequent CSI analyses will also focus on, for example, AGN and any
connections to group and/or host galaxy properties, on differences between centrals and
satellites \citep[e.g.][]{berlind2002,weinmann2006,vdb2007}, and on merger rates \citep[e.g.][]{williams2011} as functions of environment.

\section{acknowledgments}

D.D.K. expresses his appreciation to his
co-investigators for their patience. The whole team also appreciates the enormous contributions of the
Carnegie Institution for Science to the project, from the new disperser to the generosity of the
Time Assignment Committee. We are grateful to NOAO for its contributions and to the astronomical community
for its awarding of survey status to CSI in 2009. R.J.W. also gratefully acknowledges support from NSF
Grant AST-0707417. Based, in part, on observations obtained with MegaPrime/MegaCam, a joint project
of CFHT and CEA/DAPNIA, at the Canada-France-Hawaii Telescope (CFHT) which is operated by the
National Research Council (NRC) of Canada, the Institut National des Science de l'Univers of the
Centre National de la Recherche Scientifique (CNRS) of France, and the University of Hawaii. This
work is based in part on data products produced at TERAPIX and the Canadian Astronomy Data Centre as
part of the Canada-France-Hawaii Telescope Legacy Survey, a collaborative project of NRC and CNRS.
This publication makes use of data products from the Two Micron All Sky Survey, which is a joint
project of the University of Massachusetts and the Infrared Processing and Analysis
Center/California Institute of Technology, funded by the National Aeronautics and Space
Administration and the National Science Foundation

\end{document}